\documentclass{article}

\usepackage{arxiv}

\usepackage[utf8]{inputenc} % allow utf-8 input
\usepackage[T1]{fontenc}    % use 8-bit T1 fonts
\usepackage{hyperref}       % hyperlinks
\usepackage{url}            % simple URL typesetting
\usepackage{booktabs}       % professional-quality tables
\usepackage{amsfonts}       % blackboard math symbols
\usepackage{nicefrac}       % compact symbols for 1/2, etc.
\usepackage{microtype}      % microtypography
\usepackage{lipsum}
\usepackage{graphicx}
\graphicspath{ {./images/} }

\usepackage{graphicx}%
\usepackage{multirow}%
\usepackage{amsmath,amssymb,amsfonts}%
\usepackage{amsthm}%
\usepackage{mathrsfs}%
\usepackage[title]{appendix}%
\usepackage{xcolor}%
\usepackage{textcomp}%
\usepackage{manyfoot}%
\usepackage{booktabs}%
\usepackage{algorithm}%
\usepackage{algorithmicx}%
\usepackage{algpseudocode}%
\usepackage{listings}%
\usepackage{natbib}

\usepackage{xcolor}
\definecolor{gDark}{HTML}{1A1A1A}
\definecolor{gMid}{HTML}{4D4D4D}
\definecolor{gText}{HTML}{2B2B2B}
\definecolor{gFill}{HTML}{F5F5F5}
\definecolor{gLight}{HTML}{E0E0E0}
\definecolor{gRule}{HTML}{999999}
\definecolor{gArrow}{HTML}{444444}

\usepackage{tikz}
\usepackage{pgfplots}
\usepackage{pgfplotstable}
\pgfplotsset{compat=1.18}
\usetikzlibrary{patterns,positioning,calc,arrows.meta,shapes.geometric,decorations.pathreplacing,backgrounds,fit}
\usepgfplotslibrary{fillbetween}

\usepackage{algorithm, algpseudocode}
\usepackage[ruled,linesnumbered, algo2e]{algorithm2e}

\usepackage{caption}
\usepackage{subcaption}
\usepackage{float}
\usepackage{multirow}
\usepackage{siunitx}
\usepackage{colortbl}

\usepackage{threeparttable}
\usepackage{makecell}
\usepackage{cals}

\definecolor{phi}{RGB}{31,119,180}
\definecolor{deepseek}{RGB}{255,127,14}
\definecolor{opt}{RGB}{44,160,44}
\definecolor{tinyllama}{RGB}{214,39,40}
\definecolor{gridgray}{RGB}{220,220,220}

\definecolor{fedcolor}{gray}{0.35}
\definecolor{centcolor}{gray}{0.75}

\usepackage{pgfplots}
\usepgfplotslibrary{fillbetween}
\pgfplotsset{compat=1.18}
\definecolor{acccolor}{RGB}{31,119,180}
\definecolor{preccolor}{RGB}{44,160,44}
\definecolor{reccolor}{RGB}{214,39,40}
\definecolor{f1color}{RGB}{148,103,189}
\definecolor{auccolor}{RGB}{255,127,14}
\definecolor{tableheader}{RGB}{240,240,245}
\definecolor{bestcell}{RGB}{232,245,233}

\definecolor{phi}{RGB}{0,114,178}        % Blue
\definecolor{deepseek}{RGB}{230,159,0}   % Orange
\definecolor{opt}{RGB}{0,158,115}        % Green
\definecolor{tinyllama}{RGB}{204,121,167} % Pink
\definecolor{gridgray}{RGB}{220,220,220}
\definecolor{lightblue}{RGB}{235,245,251}
\definecolor{lightgreen}{RGB}{235,251,238}
\definecolor{lightorange}{RGB}{255,248,235}
\definecolor{lightpink}{RGB}{251,240,246}
\definecolor{darkred}{RGB}{180,0,0}

\definecolor{blue1}{RGB}{31,119,180}
\definecolor{orange1}{RGB}{255,127,14}
\definecolor{green1}{RGB}{44,160,44}
\definecolor{red1}{RGB}{214,39,40}
\definecolor{gray1}{RGB}{127,127,127}

\definecolor{headerblue}{RGB}{0,102,204}
\definecolor{successgreen}{RGB}{0,128,0}
\definecolor{warningorange}{RGB}{255,165,0}

\definecolor{serverblue}{RGB}{41,128,185}
\definecolor{client1}{RGB}{39,174,96}
\definecolor{client2}{RGB}{230,126,34}
\definecolor{client3}{RGB}{142,68,173}
\definecolor{dpred}{RGB}{192,57,43}
\definecolor{loracolor}{RGB}{155,89,182}
\definecolor{frozencolor}{RGB}{120,120,120}
\definecolor{inputcolor}{RGB}{52,73,94}
\definecolor{embedcolor}{RGB}{52,152,219}
\definecolor{transcolor}{RGB}{46,204,113}
\definecolor{outputcolor}{RGB}{231,76,60}
\definecolor{aggcolor}{RGB}{127,140,141}
\definecolor{bestcolor}{RGB}{39,174,96}
\definecolor{headercolor}{RGB}{41,128,185}
\definecolor{lightgray}{RGB}{245,245,245}

\theoremstyle{thmstyleone}%
\theoremstyle{thmstyletwo}%

\theoremstyle{thmstylethree}%
\newtheorem{definition}{Definition}%

\begin{document}

\title{DP-FlogTinyLLM: Differentially private federated log anomaly detection using Tiny LLMs}

\author{
 Isaiah Thompson \\
  Department of Mathematical Sciences\\
  University of Texas at El Paso\\
  El Paso, TX 79968, USA\\
  \texttt{iocansey@miners.utep.edu} \\
  %% examples of more authors
   \And
 Tanmay Sen \\
  SQC \& OR Unit\\
  Indian Statistical Institute\\
  Kolkata,  700118,   India \\
  \texttt{tanmay.sen@isical.ac.in} \\
  \And
 Ritwik Bhattacharya\\
  Department of Mathematical Sciences\\
  University of Texas at El Paso\\
El Paso, TX 79968, USA\\
  \texttt{ritwik.bhatta@gmail.com} 
}

%\begin{document}
\maketitle
\begin{abstract}
Modern distributed systems generate massive volumes of log data that are critical for detecting anomalies and cyber threats. However, in real world settings, these logs are often distributed across multiple organizations and cannot be centralized due to privacy and security constraints. Existing log anomaly detection methods, including recent large language model (LLM) based approaches, largely rely on centralized training and are not suitable for such environments. In this paper, we propose DP-FLogTinyLLM, a privacy preserving federated framework for log anomaly detection using parameter efficient LLMs. Our approach enables collaborative learning without sharing raw log data by integrating federated optimization with differential privacy. To ensure scalability in resource constrained environments, we employ low rank adaptation (LoRA) for efficient fine tuning of Tiny LLMs at each client. Empirical results on the Thunderbird and BGL datasets show that the proposed framework matches the performance of centralized LLM based methods, while incurring additional computational overhead due to privacy mechanisms. Compared to existing federated baselines, DP-FLogTinyLLM consistently achieves higher precision and F1-score, with particularly strong gains on the Thunderbird dataset, highlighting its effectiveness in detecting anomalies while minimizing false positives.
\end{abstract}
% keywords can be removed
\keywords{Differential privacy, Federated learning, Large language models, Log anomaly detection, LoRA}

\section{Introduction}\label{sec1}

Recent advancements in modern computing infrastructure have led to the generation of vast amounts of data, particularly system logs. These logs provide vital information regarding system behavior, ranging from routine system activities to system failures  and intrusions. System failures  and instructions are anomalous events found in system logs, triggering concerns about system reliability and security \cite{chourasiya2025advanced}. Over the past decade, there have been models ranging from rule-based pattern recognition to deep learning sequential architectures to detect anomalies in system logs as a language modeling task \cite{mdpi2025logstudy}.  DeepLog \cite{du2017deeplog} demonstrated the use of Long Short Term Memory (LSTM) to predict the next log key sequence and flags deviations as anomalies. LogAnomaly \cite{meng2019loganomaly} extended the LSTM paradigm to detect both sequential and quantitative log anomalies concurrently by modeling high-dimensional dependencies within template sequences and flagging observed template sequences as anomalies if they are not among the top k predictions. More recently, transformer-based models such as  LogBERT \cite{guo2021logbert}, LogGPT \cite{han2023loggpt} , LogTinyLLM \cite{ocansey2025logtinyllm}, among others, have gained significant improvements in log anomaly detection.

However, every one of these models assumes that all log data is available at a single location for centralized training. In practice, this assumption rarely holds, since log data may be produced by servers in different geographic locations and data centers \cite{arxiv2512_08277}. Regulations such as the European Union's General Data Protection Regulation \cite{gdpr2016info} and China's Personal Information Protection Law \cite{zhu2022pipl} treat system logs as sensitive data because they contain behavioral traces, user session identifiers, and other personal information. Organizations operating in different geographical locations need large amounts of data to train a robust log anomaly detection model, but may face legal challenges in transmitting this data to a centralized location due to differing legal compliance requirements across these locations \cite{nist_ppfl_pipeline_2024}. The consequence is that most state-of-the-art models that assume centralized data cannot be deployed by some of these organizations that operate in different geographical locations with different data protection laws \cite{jmir2023_e41588}. The federated learning log anomaly detction paradigm offers a seamless solution to this problem by allowing multiple servers at different geographical locations to train a shared model while keeping their raw data local \cite{vucovich2022anomaly}. Despite this solution that federated learning offers, there are only a few studies in this area for log anomaly detection \cite{bithi2026adaptive}. The study \cite{landauer2024federated} was among the first to explore this area by applying a federated LSTM architecture to HDFS log data. However, their approach did not use any language model and did not provide any formal privacy guarantees. \\

Studies such as \cite{shin2023utility, tsouvalas2025enccluster} provide initial baselines for federated log anomaly detection. However, it remains unclear whether these approaches can achieve performance comparable to centralized language model based methods. Extending existing LLM-based models to federated settings introduces an additional challenge related to computational cost and memory requirements. For instance, LogLLM \cite{guan2024logllm} combines BERT-base with LLaMA-3-8B, requiring approximately 16 GB of GPU memory for inference alone and significantly more for training \cite{bentoml2024gpu_memory_llms}. Techniques such as LoRA \cite{hu2022lora} and quantization can reduce the computational burden to some extent. However, these methods alone are often insufficient for resource-constrained environments. Therefore, the use of compact (tiny) language models becomes essential for enabling efficient training while maintaining reasonable performance \cite{ocansey2025logtinyllm}.\\

The federated learning log anomaly detection models keep the raw data local, but the model updates are shared at each round  for aggregation, which are not inherently private. Research on gradient inversion has shown that individual training examples can be reconstructed from transmitted gradients with surprising fidelity \cite{eltaras2025rconvpp}. The standard guard against this is the use of a differential privacy algorithm introduced by \cite{abadi2016deep}, which clips per-sample gradients to a fixed norm and adds calibrated Gaussian noise before each update. This provides a formal, mathematical guarantee  expressed as an  $(\varepsilon, \delta)$ budget  on the maximum information any adversary can extract about any single training example \cite{abadi2016deep}. Despite its importance in federated log anomaly detection, the general federated DP literature has focused on image classification and general NLP tasks \cite{xu2025dpfedlora}, rather than on the sequential anomaly-detection objective that log models optimize. Another issue in a federated learning log anomaly detection setting is heterogeneity in the log data across servers at different geographical locations, which can lead to a non-IID structure. This is a precise setting where standard federated averaging struggles \cite{mcmahan2017communication,alenezi2024iot_intrusion} .  The study \cite{li2020fedprox} demonstrated that FedAvg can diverge when local data distributions are heterogeneous, and proposed FedProx, which adds a proximal regularization term to each client's local objective to limit drift from the global model. Putting all together, Centralized LLM-based methods such as LogGPT and LogLLM maximize utility but provide no privacy \cite{han2023loggpt}. In federated log anomlay detection literature, no study has explored tiny  language models, Differential privacy, and FedProx concurrently.\\

 In this paper, we propose DP-FlogTinyLLM, referred to simply as FlogTinyLLM, a federated framework that addresses privacy and heterogeneity issues in system logs for log anomaly detection. FlogTinyLLM is built around  four distinct tiny LLMs, enhanced with low-rank adaptation, that are small enough to train entirely on edge hardware yet expressive enough to capture the sequential patterns of parsed log keys. Training uses four distinct compact sized language models: Microsoft Phi-1.5, DeepSeek-R1-Qwen, Facebook OPT-1.3, and TinyLlama-1.1B with FedProx aggregation incorporating differential privacy accounting, providing a tracked $(\varepsilon, \delta)$  privacy budget across all communication rounds. To our knowledge, this is the first framework that unifies federated learning, differential privacy, FedProx, and edge-scale models for log anomaly detection. The main contributions of this work are as follows:
\begin{itemize}
	
	\item We leverage four distinct tiny large language models with LoRA adaptation, enabling fully on-device federated training within edge memory constraints. This contrasts with existing LLM-based log anomaly detection approaches that typically require models with 8B+ parameters and over 30 GB of GPU memory.
	
	\item We incorporated DP-SGD with R\'{e}nyi differential privacy accounting into the federated training pipeline, providing formal $(\varepsilon, \delta)$-differential privacy guarantees with a fully tracked privacy budget. 
	
	\item Through comprehensive evaluation on the Thunderbird and BGL log datasets, we demonstrate that DP-FlogTinyLLM achieves superior performance compared to existing federated benchmark models for log anomaly detection.
	
\end{itemize}

\subsection{Related Work}

Detecting log anomalies in distributed systems by allowing multiple clients to train a shared model  while keeping each client's data local has attracted significant reserach attention. Several federated log anomaly detection models have been proposed over the past decades, each with a distinct underlying architecture \cite{du2022deeplog}. Fedlog \cite{li2022federated} combines a Temporal Convolutionary Network with an attention-based convolutional network (TCN-ACNN) with a softmax layer to classify log sequences as normal or anomalous. While the model is trained across clients in an unsupervised manner with the capability of capturing temporal patterns in log sequences, it does not belong to the family of language models that have proven effectiveness for sequential log analysis. The same observation applies to FLOGCNN ~\cite{guo2021anomaly}. FLOGCNN is a lightweight model that uses a standard neural network and federated averaging (FedAvg) as its aggregating strategy in each communication round.  The model is designed to extract inherent strategies in log sequences, but it lacks the capability to model long-range dependencies in log sequences, which is a problem that transformer-based models solved in centralized settings. Shin et al.~\cite{shin2023utility} took a different route by evaluating two well-established deep learning architectures within a federated framework: one-dimensional convolutional neural networks (CNN1D) and Long Short-Term Memory (LSTM) networks. Their approach shows that classical models can be adopted in a federated setting, but it does not introduce architectural innovations pertinent to sequential log anomaly detection. Federated deeplog ~\cite{himler2024anomaly} directly focused on cyber security applications by adapting a semi-supervised LSTM model from the open-sourced LogDeep framework~\cite{10}. Their approach leverages the Flower framework to manage the federated loop. Among the existing federated log anomlay detection architectures, Federated deeplog responds to the practical challenges of the problem but remains underpinned by LSTM, which trails the effectiveness of language model based methods in centralized evaluations. Most of the existing architectures rely on TCNN, CNN, or LSTM architectures that predate the transformer revolution in sequential modeling. Strongest centralized results on standard log benchmarks now come from language model based approaches that treat log key prediction as a natural language modeling task \cite{zhou2024llm_log_parsing}. FlogTinyLLM addresses this directly by adopting  four distinct transformer based tiny language models enhanced with low-rank adaptation, bridging language-level sequential understanding into federated learning.\\

Another limitation of existing work is the lack of proper consideration of privacy. Although federated learning keeps client data local, it does not protect against adversarial attacks. Model updates shared during communication rounds are not inherently private and may leak information about the underlying data. Among the existing models, none provides rigorous privacy guarantees \cite{liu2026federatedhealth}. Fedlog ~\cite{li2022federated} relies on the federated learning paradigm itself as the basis for their privacy claims, with no privacy budget computed, no noise is added to model updates, and no analysis offered to quantify how much information the shared parameters might reveal about any individual client's data. FLOGCNN ~\cite{guo2021anomaly} made mention of homomorphic encryption as a potential mechanism for protecting local parameters, but this remains at the level of assumption. No encryption scheme is actually implemented or evaluated within the training pipeline~\cite{shin2023utility}, leaving the system exposed to information leakage through the model updates shared during each training round. Federated Deeplog~\cite{himler2024anomaly} acknowledges data privacy as a concern and frames it as a motivating challenge, but their implementation does not include differential privacy or any other formal mechanism to bound the information that aggregated updates reveal.\\

FlogTinyLLM addresses this challenge by integrating differential privacy directly into the federated training process. Each client clips its local gradients and adds calibrated noise before transmitting updates to the server, following the DP-SGD framework. The resulting privacy cost is tracked across all communication rounds using R\'{e}nyi differential privacy accounting, producing a concrete $(\varepsilon, \delta)$ budget that quantifies the maximum information leakage.  \\

Another limitation of existing methods is their inability to handle differences in data distributions across clients. In real settings, each client has its own workload and operating conditions, leading to different patterns of normal behavior and anomalies. This results in non-IID data, which is common in federated log settings \cite{zhang2024anomaly_fl_review}. Despite this, the aggregation strategies used in existing work do not account for it. FLOGCNN entirely avoids the problem  by partitioning the training data into 50 equal sized clients, creating an artificially uniform distribution that does not reflect what would occur in practice ~\cite{guo2021anomaly}. Deeplog(FL) ~\cite{himler2024anomaly} uses FedAvg for model aggregation, which assumes that client data is approximately identically distributed. When this assumption is violated as it inevitably is with real log data, FedAvg can produce a global model that converges slowly, converges to a poor solution, or fails to converge altogether \cite{li2019fedavg_noniid}. \\

FlogTinyLLM addresses data heterogeneity at two levels. At the data level, it assigns compute nodes derived from the training data to 14 clients on the Thunderbird dataset and 15 clients on the BGL dataset using a Round Robin algorithm. This creates a federated training environment in which each client holds a distinct subset of the data that may contain different patterns of normal behavior and different types of anomalies, reflecting the conditions of a real world multi-site setting. At the algorithmic level, FlogTinyLLM replaces FedAvg with FedProx~\cite{li2020fedprox}. FedProx adds a proximal penalty term to each client's local loss function, discouraging the local model from drifting too far from the global model during training. This penalty ensures that even when individual clients see very different data distributions, their local updates remain compatible with the shared global objective, promoting stable convergence across heterogeneous clients.

\section{Problem Statement}

We consider a federated setting with $N$ clients, where each client $C_i$ holds a private dataset $\mathcal{D}_i$ consisting of log key sequences. The data remains local to each client, and raw logs are not shared with the central server or other clients. The goal is to learn a global model $W$ that can detect anomalies in log sequences across all clients. Each client minimizes a local loss function $\mathcal{L}_i(W)$ based on its own data. The standard federated learning objective is given by:

\begin{equation}
	\min_{W} \sum_{i=1}^{N} \frac{|\mathcal{D}_i|}{|\mathcal{D}|} \mathcal{L}_i(W),
\end{equation}

\noindent where $|\mathcal{D}_i|$ is the number of samples on client $C_i$, and $|\mathcal{D}| = \sum_{i=1}^{N} |\mathcal{D}_i|$ is the total number of samples across all clients. To handle data heterogeneity, we adopt the FedProx formulation \cite{li2020fedprox}, where each client solves the following local objective at communication round $t$:

\begin{equation}
	\min_{W} \ \mathcal{L}_i(W) + \frac{\mu}{2} \|W - W_t\|^2,
\end{equation}

\noindent where $W_t$ denotes the global model at round $t$, and $\mu$ controls the strength of the proximal term. In addition, the training process must satisfy $(\varepsilon, \delta)$-differential privacy. This ensures that for any two neighboring datasets differing in a single training example, the probability of any output of the algorithm differs by at most a factor of $e^{\varepsilon}$, up to a small probability $\delta$ \cite{apxml_differential_privacy_fl}. This constraint limits the amount of information that can be inferred about any individual data point from the shared model updates.

\section{Methodology}
% This section outlines the detailed step-by-step methodology employed in this study.
This section describes the FlogTinyLLM pipeline. We begin by formally stating the problem, then walk through each stage from raw log messages to anomaly detection. 

\subsection{Log Parsing }
\begin{figure*}[h]
	\centering
	\resizebox{\textwidth}{!}{%
		\begin{tikzpicture}[
			>=Stealth,
			every node/.style={font=\sffamily},
			arr/.style={-{Stealth[length=5pt,width=4pt]}, 
				line width=0.8pt, color=gArrow}
			]
			
			% ???????????????????????????????????????????????
			%  STAGE 1: RAW LOG LINES
			% ???????????????????????????????????????????????
			
			% Header
			\node[fill=gDark, text=white, rounded corners=2pt,
			minimum height=0.5cm, minimum width=14cm,
			font=\sffamily\fontsize{9}{11}\selectfont\bfseries,
			align=center] (rawheader) at (0, 0) 
			{Raw Thunderbird Syslog Entries};
			
			% Raw log content box
			\node[draw=gDark, fill=white, rounded corners=2pt,
			text width=13.6cm, inner sep=8pt, line width=0.5pt,
			below=0.15cm of rawheader, align=left,
			font=\ttfamily\fontsize{7}{9.5}\selectfont, 
			text=gText] (rawbox) {%
				\textbf{1131566461 2005.11.09 dn228} Nov 9 12:01:01 dn228/dn228 crond(pam\_unix)[2915]: session closed for user root\\[3pt]
				\textbf{1131566461 2005.11.09 dn228} Nov 9 12:01:01 dn228/dn228 crond(pam\_unix)[2915]: session opened for user root by (uid=0)\\[3pt]
				\textbf{1131566461 2005.11.09 dn228} Nov 9 12:01:01 dn228/dn228 crond[2916]: (root) CMD (run-parts /etc/cron.hourly)\\[3pt]
				\textbf{1131566461 2005.11.09 dn261} Nov 9 12:01:01 dn261/dn261 crond(pam\_unix)[2907]: session closed for user root\\[3pt]
				\textbf{1131566461 2005.11.09 dn261} Nov 9 12:01:01 dn261/dn261 crond(pam\_unix)[2907]: session opened for user root by (uid=0)};
			
			% Annotation: what each field is
			\node[font=\sffamily\fontsize{6}{7.5}\selectfont, 
			text=gMid, anchor=north west, align=left]
			(rawfooter) at ([yshift=-0.15cm]rawbox.south west) {%
				\textit{Fields:}~~ 
				Unix Timestamp~\textbar~~Date~\textbar~~
				\textbf{Compute Node}~\textbar~~
				Human Date/Time~\textbar~~
				Node/Node~\textbar~~Component[PID]:~\textbar~~
				Message Content};
			
			% ???????????????????????????????????????????????
			%  ARROW 1: Raw ? Parser
			% ???????????????????????????????????????????????
			
			\node[draw=gDark, fill=gLight, rounded corners=3pt,
			minimum width=3.5cm, minimum height=0.65cm,
			font=\sffamily\fontsize{8}{10}\selectfont\bfseries,
			text=gDark, align=center, line width=0.6pt,
			below=0.9cm of rawfooter] (drainbox) 
			{Drain Parser};
			
			% \node[font=\sffamily\fontsize{7}{9}\selectfont, 
			%       text=gMid, right=0.4cm of drainbox, align=left] {%
				%       st $= 0.3$,~~depth $= 3$\\
				%       Regex on content field only};
			
			\draw[arr] ([yshift=-0.05cm]rawfooter.south) -- (drainbox.north);
			
			% ???????????????????????????????????????????????
			%  STAGE 2: PARSED TEMPLATES WITH NODE INFO
			% ???????????????????????????????????????????????
			
			% Header
			\node[fill=gDark, text=white, rounded corners=2pt,
			minimum height=0.5cm, minimum width=14cm,
			font=\sffamily\fontsize{9}{11}\selectfont\bfseries,
			align=center, below=0.9cm of drainbox] (parsedheader) 
			{Parsed Log};
			
			\draw[arr] (drainbox.south) -- (parsedheader.north);
			
			% Column headers for parsed output
			\node[font=\sffamily\fontsize{7}{9}\selectfont\bfseries,
			text=gDark, below=0.2cm of parsedheader, 
			anchor=north west, xshift=-6.6cm] (colhdr) {%
				\begin{tabular}{@{}
						p{0.8cm}@{\hspace{6pt}}
						p{1.8cm}@{\hspace{6pt}}
						p{1.2cm}@{\hspace{6pt}}
						p{8.2cm}@{}}
					\textbf{Event} & \textbf{Timestamp} & 
					\textbf{Node} & \textbf{Template} \\
			\end{tabular}};
			
			% Separator under column headers
			\draw[gDark, line width=0.4pt] 
			([yshift=-0.05cm]colhdr.south west) -- 
			++(13.6cm, 0);
			
			% Parsed entries
			\node[draw=gDark, fill=gFill, rounded corners=2pt,
			text width=13.6cm, inner sep=8pt, line width=0.5pt,
			below=0.08cm of colhdr.south west, anchor=north west,
			align=left] (parsedbox) {%
				\begin{tabular}{@{}
						p{0.8cm}@{\hspace{6pt}}
						p{1.8cm}@{\hspace{6pt}}
						p{1.2cm}@{\hspace{6pt}}
						p{8.2cm}@{}}
					{\sffamily\fontsize{7.5}{10}\selectfont\bfseries $E_1$} &
					{\ttfamily\fontsize{6.8}{9}\selectfont 1131566461} &
					{\ttfamily\fontsize{6.8}{9}\selectfont\bfseries dn228} &
					{\ttfamily\fontsize{6.8}{9}\selectfont 
						crond(pam\_unix)[<*>]: session closed for user <*>} 
					\\[5pt]
					{\sffamily\fontsize{7.5}{10}\selectfont\bfseries $E_2$} &
					{\ttfamily\fontsize{6.8}{9}\selectfont 1131566461} &
					{\ttfamily\fontsize{6.8}{9}\selectfont\bfseries dn228} &
					{\ttfamily\fontsize{6.8}{9}\selectfont 
						crond(pam\_unix)[<*>]: session opened for user <*> by (uid=<*>)} 
					\\[5pt]
					{\sffamily\fontsize{7.5}{10}\selectfont\bfseries $E_3$} &
					{\ttfamily\fontsize{6.8}{9}\selectfont 1131566461} &
					{\ttfamily\fontsize{6.8}{9}\selectfont\bfseries dn228} &
					{\ttfamily\fontsize{6.8}{9}\selectfont 
						crond[<*>]: (<*>) CMD (run-parts /etc/cron.hourly)} 
					\\[5pt]
					{\sffamily\fontsize{7.5}{10}\selectfont\bfseries $E_1$} &
					{\ttfamily\fontsize{6.8}{9}\selectfont 1131566461} &
					{\ttfamily\fontsize{6.8}{9}\selectfont\bfseries dn261} &
					{\ttfamily\fontsize{6.8}{9}\selectfont 
						crond(pam\_unix)[<*>]: session closed for user <*>} 
					\\[5pt]
					{\sffamily\fontsize{7.5}{10}\selectfont\bfseries $E_2$} &
					{\ttfamily\fontsize{6.8}{9}\selectfont 1131566461} &
					{\ttfamily\fontsize{6.8}{9}\selectfont\bfseries dn261} &
					{\ttfamily\fontsize{6.8}{9}\selectfont 
						crond(pam\_unix)[<*>]: session opened for user <*> by (uid=<*>)} 
					\\
			\end{tabular}};
			
			% % Annotation about wildcards and node preservation
			% \node[font=\sffamily\fontsize{6.5}{8}\selectfont, 
			%       text=gMid, anchor=north west, align=left]
			%       (parsedfooter) at 
			%       ([yshift=-0.15cm]parsedbox.south west) {%
				% \textit{Key:}~~\texttt{<*>} replaces 
				% variable fields (PIDs, usernames, UIDs).~~
				% \textbf{Compute node preserved} for 
				% downstream sliding window grouping.~~
				% Same template $\to$ same Event ID 
				% regardless of node.};
			
			% %
		\end{tikzpicture}
	}% end resizebox
	\caption{shows parsed Thunderbird log data. Raw syslog entries are processed by the Drain algorithm, which extracts event templates by replacing variable fields with wildcards while preserving the originating compute node.}
	\label{fig:thunderbird_log_parsing}
\end{figure*}
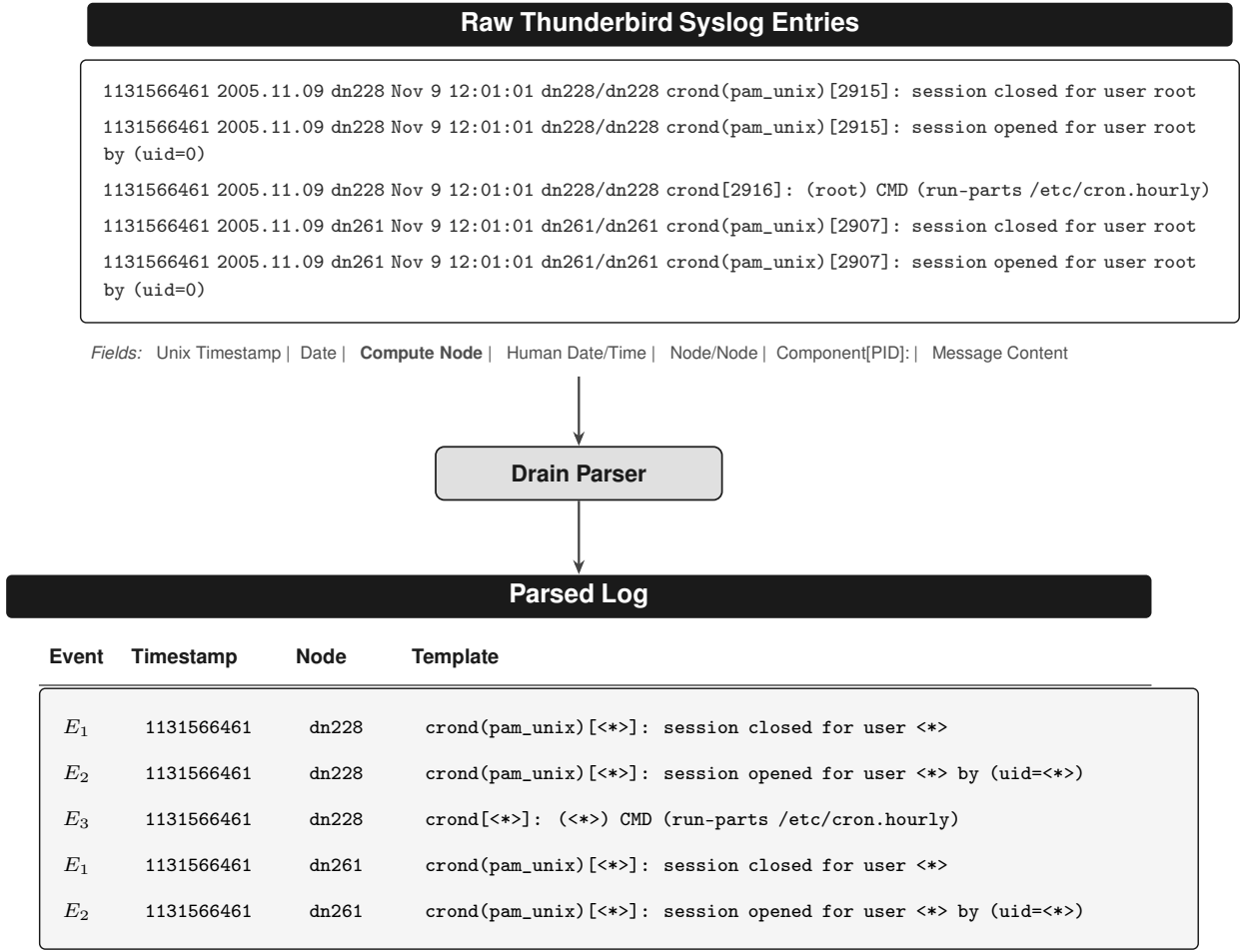

Raw log messages are unstructured or semi-structured and are converted to a structured representation before being used as input to a language model. FlogTinyLLM performs this conversion using the Drain algorithm~\cite{he2017drain}. Drain organizes incoming log messages into a fixed-depth tree based on textual similarity. Each leaf of this tree corresponds to a distinct message template. Variable fields such as hostnames and process identifiers within these templates are replaced with wildcards ~\cite{he2017drain}. This process is called \textit{Log parsing}. Figure \ref{fig:thunderbird_log_parsing} illustrates an example of the parsed Thunderbird log data

\subsection{Sliding Window Construction}

Following the approach established by\cite{han2023loggpt}, the parsed log entries are grouped into sliding time windows to form a sequence of log keys. Each window spans 5 minutes, and consecutive windows overlap by a 1-minute step size. A window is labeled as anomalous if it contains at least one log entry that is marked as anomalous in the original log dataset. Otherwise, the window is labeled as normal.

\subsection{Round-Robin Distribution}
To simulate a realistic multi-site setting, sequences of the log keys are distributed by their compute nodes to N clients using Round Robin algorithm \cite{silberschatz2018os}. The Thunderbird dataset contains approximately 9,024 compute nodes \cite{oliner2007supercomputers}, while the BGL dataset contains about 65,536 compute nodes \cite{almasi2003mpi}. The Round Robin assignment works as follows: the first compute node is assigned to client 1, the second to client 2, the third to client 3, and so on up to client $N$, after which the cycle repeats. This produces a data distribution where each client holds a distinct subset of the log dataset.

\subsection{ Model Architecture }
FlogTinyLLM adopted a horizontal federated learning (HFL) approach because all participating compute nodes distributed to clients originate from the same set of supercomputers. This means that while each client operates on data that may be unique in certain aspects, their overall feature space remains the same. FlogTinyLLM uses  the sequence of log keys for each client as  input to the following tiny language models enhanced with low-rank adaptation:

\paragraph{OPT-1.3B:} Zhang et al.~\cite{zhang2022opt} introduced open pre-trained transformers (OPT), a family of auto-regressive causal language models (decoder only) developed by Meta AI, which range from 125 million to 175 billion parameters, and are open source. OPT-1.3B is a mid-sized model within this family, which demonstrates competitive performance compared to models of similar scale. The open access nature of OPT-1.3B allows for flexible integration and tuning, facilitating domain-specific adaptation without the high costs associated with training a model from scratch. The model supports a context window suitable for modeling sequential dependencies in log data and provides reliable inference performance for log classification. Opt 1.3B consists of 24 layers, each containing 32 attention heads, an embedding size of 2048, and a vocabulary size of 50272 \cite{ocansey2025logtinyllm}.

\paragraph{Phi-1.5: } Li et al.~\cite{li2023textbooks} introduced Phi-1.5, a decoder only transformer based language model with 1.3 billion parameters developed by Microsoft and designed for efficient reasoning, coding, and math problem-solving. It was trained on a high quality dataset consisting of 30 billion tokens, which includes 7 billion tokens from Phi-1's filtered code corpus and 20 billion synthetically generated tokens modeled after textbooks. The model has 24 transformer layers, each with 32 attention heads and a hidden size of 2048, using Rotary Positional Embeddings (RoPE) with a rotary dimension of 32. It uses FlashAttention and a code friendly tokenizer, making it ideal for real time, low resource applications without the need for instruction tuning or RLHF. Despite its small size, Phi-1.5 performs competitively with much larger models like LLaMA-2 7B, while being more lightweight and resource efficient \cite{ocansey2025logtinyllm}. \\

\paragraph{TinyLlama-1.1B:} Zhang et al.~\cite{zhang2024tinyllama} introduced TinyLlama, an open source language model with 1.1 billion parameters based on the LLaMA 2 architecture. It is built as a decoder-only transformer with 22 transformer layers, 32 attention heads, and a hidden size of 2048. The latest version, TinyLlama v1.1, was trained on up to 2 trillion tokens, utilizing a multi-stage pre-training process that incorporated domain-specific corpora, including SlimPajama, StarCoder, ProofPile, and Skypile. This paper utilizes TinyLlama-1.1B due to its excellent performance-to-size ratio and efficient architecture. With a context window of 2,048 tokens and optimized features such as Rotary Positional Embeddings, Grouped-Query Attention, SwiGLU activations, and FlashAttention-2, it provides fast and accurate detection of anomalies in log sequences\cite{ocansey2025logtinyllm}.

\paragraph{DeepSeek-R1-Distill-Qwen-1.5B:} Guo et al.~\cite{guo2025deepseek} introduced DeepSeek-R1, a family of first generation reasoning optimized language models developed using reinforcement learning (RL) without supervised fine tuning (SFT) as a preliminary step. DeepSeek-R1-Distill-Qwen-1.5B model is distilled from the larger DeepSeek-R1 model, with the base architecture of Qwen-2.5-Math-1.5B. It consists of 28 transformer layers, each with a hidden size of 2048, and uses Grouped-Query Attention (GQA), where 32 attention heads are split into 12 query heads and 2 shared key value heads, reducing memory cost while preserving performance. The model employs RoPE to encode position information within queries and keys, supporting a maximum context length of 32,768 tokens. It uses SwiGLU activation in the feedforward blocks for better non-linearity. The feedforward network has an intermediate size of 5632. This paper utilizes DeepSeek-R1-Distill-Qwen-1.5B due to its efficient architecture and proven performance in reasoning tasks. The model benefits from distilled reasoning behaviors inherited from larger base models, allowing it to process structured log sequences effectively \cite{ocansey2025logtinyllm}.

\subsection{Training with Low-Rank Adaptation (LoRA)}

Training full scale large language models on each client is often infeasible due to limited computational resources. To overcome this, FlogTinyLLM employs Low-Rank Adaptation \cite{hu2022lora}, which enables efficient fine-tuning with a small number of additional parameters on the tiny language models outlined above. Given a weight matrix $W \in \mathbb{R}^{d \times d}$, LoRA models the update as:
\begin{equation}
	W_{\text{adapted}} = W + BA,
\end{equation}
where $B \in \mathbb{R}^{d \times r}$ and $A \in \mathbb{R}^{r \times d}$ with $r \ll d$. This significantly reduces the number of trainable parameters, making LLM-based anomaly detection practical in federated environments. In transformer models, LoRA is applied to the query, key, and value projection matrices. The attention mechanism is defined as:
\begin{equation}
	\text{Attention}(Q, K, V) = \text{softmax} \left( \frac{QK^\top}{\sqrt{d_k}} \right)V,
\end{equation}
where $Q = HW_Q$, $K = HW_K$, and $V = HW_V$.

Consider each log sequence is represented as $(k_1, k_2, \dots, k_T)$, where $k_t$ denotes the log key at time step $t$. In particular, the embedding layers of each of these tiny language models convert each log key $k_i$ into a dense vector $\mathbf{e}_i \in \mathbb{R}^{d}$, where $d$ is the hidden dimension of the tiny language model. A learned positional embedding is added to each $\mathbf{e}_i$  to preserve the positions of each of the log keys within the sequence. The result is a matrix of input representations $[\mathbf{e}_1, \mathbf{e}_2, \ldots, \mathbf{e}_T]$ that serves as input to the transformer blocks of the language models.  Inside each transformer layer, the input representations are projected into queries (Q), keys (K), and values (V) through weight matrices $W_Q$, $W_K$, and $W_V$. In a standard transformer~\cite{vaswani2017attention}, these projections take the form:  \begin{equation} Q = \mathbf{H} W_Q, \quad K = \mathbf{H} W_K, \quad V = \mathbf{H} W_V \label{eq:standard_proj} \end{equation}  \noindent where $\mathbf{H} = [\mathbf{e}_1, \ldots, \mathbf{e}_T]$ is the matrix of input embeddings. These projections determine what each log key attends to within the sequence. The attention output is then computed as:  \begin{equation} \text{Attention}(Q, K, V) = \text{softmax}\!\left( \frac{QK^\top}{\sqrt{d_k}}\right) V \label{eq:attention} \end{equation}  \noindent where $d_k$ is the dimension of each attention head. LoRA keeps all three of these weight matrices frozen at their pretrained values. Rather than updating them, LoRA attaches a low-rank bypass to each one following the framework~\cite{hu2022lora}. For any frozen matrix $W \in \mathbb{R}^{d \times d}$, the adapted version is:  \begin{equation} W_{adapted} = W + \frac{\alpha}{r} \, BA \label{eq:lora_decomp} \end{equation}  \noindent where $B \in \mathbb{R}^{d \times r}$ and $A \in \mathbb{R}^{r \times d}$ are small trainable matrices, $r$ is the rank of the decomposition, and $\alpha$ is a scaling factor that controls how strongly the adaptation influences the output. With this modification, the projections that produce queries, keys, and values from the log key embeddings become:  \begin{equation} Q = \mathbf{H}\!\left(W_Q + \frac{\alpha}{r} B_Q A_Q\right), \quad K = \mathbf{H}\!\left(W_K + \frac{\alpha}{r} B_K A_K\right), \quad V = \mathbf{H}\!\left(W_V + \frac{\alpha}{r} B_V A_V\right) \label{eq:lora_qkv} \end{equation}  \noindent The attention computation in Equation~\ref{eq:attention} receives the adapted $Q$, $K$, and $V$ and proceeds as usual.

The practical effect of this design on the log key embeddings is as follows. The frozen matrices $W_Q$, $W_K$, and $W_V$ already know how to project token embeddings into an attention space, because they were learned during pretraining \cite{envisioning_attention_projection_matrix}.  The LoRA matrices $A$ and $B$ learn a small correction on top of these projections that is specific to the anomaly detection task \cite{emergentmind_lora_det}. Through this correction, the model adjusts which log keys it considers relevant to each other within a sequence, without discarding the general sequential knowledge captured during pretraining\cite{hu2021lora} .  After passing through all transformer layers, the model outputs contextual representations $[h_1, h_2, \ldots, h_T]$ that reflect not only the identity of each log key but also its relationship to every other key in the sequence. At each communication round, only the LoRA parameters are communicated, with the FedProx penalty applied.  While LoRA improves computational efficiency, it does not by itself guarantee privacy. To provide formal privacy protection, we incorporate differential privacy into the federated training process.

\subsection{Federated Learning and Differential Privacy}

FlogTinyLLM operates in a federated learning setting, where training proceeds over multiple communication rounds between a central server and distributed clients, as illustrated in Fig.~2. At each round $t$, the server broadcasts the current global model $w^{(t)}$ to all participating clients. Each client then performs local training on its private log data and computes an update based on its local objective. These updates are subsequently sent back to the server for aggregation.

Although raw log data never leaves a client, the transmitted model updates can still leak information about individual log sequences. An adversary observing the global model across multiple rounds may infer whether a specific log sequence was part of a client's training data. This type of attack is known as a membership inference attack \cite{zhang2026unified_defense_fl}. To mitigate this risk, FlogTinyLLM incorporates differential privacy (DP) into the federated training process. Differential privacy, introduced by Dwork et al.~\cite{dwork2010boosting}, provides a formal guarantee that the output of a computation does not depend significantly on any single data record. The formal definition is as follows \cite{census_differential_privacy_2023}.

\begin{definition}[$(\varepsilon, \delta)$-Differential Privacy]
	A randomized mechanism $\mathcal{M}$ satisfies $(\varepsilon, \delta)$-differential privacy if, for any two datasets $D$ and $D'$ that differ in exactly one record, and for all measurable sets $S \subseteq \mathrm{Range}(\mathcal{M})$:
	\begin{equation}
		\Pr[\mathcal{M}(D) \in S] \leq 
		e^{\varepsilon} \cdot \Pr[\mathcal{M}(D') \in S] + \delta.
	\end{equation}
\end{definition}

\noindent In the context of federated learning, the mechanism $\mathcal{M}$ corresponds to the process that maps local datasets to the shared model updates and the final global model. If DP is satisfied, the inclusion or exclusion of any single log sequence in a client's dataset has only a limited effect on the distribution of the observed model outputs. In FlogTinyLLM, differential privacy is enforced during each communication round. After local training, each client computes its model update $\Delta_i$ and applies gradient clipping to bound its sensitivity:
\begin{equation}
	\|\Delta_i\|_2 \leq C.
\end{equation}

\noindent The server then aggregates the clipped updates and adds Gaussian noise:
\begin{equation}
	w^{(t+1)} = \sum_{i=1}^{K} \frac{n_i}{n} w_i + \mathcal{N}(0, \sigma^2 C^2 I),
\end{equation}

\noindent where $\sigma$ is the noise multiplier. This procedure ensures that each round contributes a controlled amount to the total privacy loss. It is note that we adopt a central differential privacy (CDP) setting, where the server is assumed to be trusted and applies Gaussian noise during aggregation.

The overall privacy guarantee is tracked across all communication rounds using R\'{e}nyi differential privacy accounting, resulting in a final $(\varepsilon, \delta)$ privacy budget. A smaller value of $\varepsilon$ corresponds to stronger privacy, meaning that the model behaves similarly regardless of whether any specific log sequence is included in training.

\paragraph{Handling Data Heterogeneity.}
To handle data heterogeneity for each client, FlogTinyLLM adopts the Federated Proximal (FedProx) algorithm \cite{li2020federated}, which extends the standard Federated Averaging (FedAvg) objective. The FedAvg objective is:
\begin{equation}
	\min_{W} \sum_{i=1}^{N} \frac{|\mathcal{D}_i|}{|\mathcal{D}|} L_i(W),
\end{equation}
FedProx introduces a proximal term:
\begin{equation}
	\min_{W} \sum_{i=1}^{N} \frac{|\mathcal{D}_i|}{|\mathcal{D}|} L_i(W) + \frac{\mu}{2} \|W - W_t\|^2,
\end{equation}
where $W_t$ is the global model at round $t$. This term constrains local updates to remain close to the global model, improving stability under non-IID  log data.

\begin{figure}[h]
	\centering
	\resizebox{\textwidth}{!}{%
		\begin{tikzpicture}[
			>=Stealth,
			box/.style={
				rectangle, 
				rounded corners=3pt, 
				draw=black, 
				fill=gray!8,
				line width=0.8pt, 
				minimum width=4.2cm, 
				minimum height=0.5cm,
				font=\footnotesize,
				align=center
			},
			serverbox/.style={
				rectangle, 
				rounded corners=4pt, 
				draw=black, 
				fill=gray!20,
				line width=1pt, 
				minimum width=4.5cm, 
				minimum height=0.7cm,
				font=\footnotesize\bfseries
			},
			dpbox/.style={
				rectangle, 
				rounded corners=3pt, 
				draw=black, 
				fill=gray!15,
				line width=0.8pt, 
				minimum width=3.8cm, 
				minimum height=0.5cm,
				font=\footnotesize
			},
			clientlabel/.style={
				font=\footnotesize\bfseries
			},
			arrow/.style={->, line width=0.7pt, black},
			bigarrow/.style={->, line width=1pt, black}
			]
			
			%==============================================================================
			% SERVER - TOP
			%==============================================================================
			\node[serverbox] (server) at (0, 9) {Server: Global Model $w^{(t)}$};
			
			%==============================================================================
			% BROADCAST
			%==============================================================================
			\draw[bigarrow] (server.south) -- ++(0, -0.4) coordinate (bc);
			\draw[bigarrow] (bc) -- ++(-5.2, 0) |- (-5.2, 7.8);
			\draw[bigarrow] (bc) -- (0, 7.8);
			\draw[bigarrow] (bc) -- ++(5.2, 0) |- (5.2, 7.8);
			
			\node[font=\scriptsize, fill=white] at (0, 8.3) {Broadcast LoRA parameters};
			
			%==============================================================================
			% CLIENT 1
			%==============================================================================
			\begin{scope}[shift={(-5.2, 0)}]
				\node[draw=black, fill=white, rounded corners=5pt, line width=1pt,
				minimum width=4.6cm, minimum height=7.2cm] (c1box) at (0, 3.8) {};
				\node[clientlabel] at (0, 7.2) {Client 1};
				\node[font=\scriptsize, gray] at (0, 6.8) {$\mathcal{D}_1$, $n_1$ samples};
				
				\node[box] (in1) at (0, 6.2) {$[k_1, k_2, \ldots, k_T]$};
				
				\node[box] (emb1) at (0, 5.4) {Embedding: $k_i \rightarrow \mathbf{e}_i$};
				\draw[arrow] (in1) -- (emb1);
				
				\node[draw=black, fill=gray!5, rounded corners=4pt, line width=0.8pt,
				minimum width=4.3cm, minimum height=1.6cm] (trans1) at (0, 4) {};
				\node[font=\scriptsize\bfseries] at (0, 4.5) {Transformer + LoRA};
				
				\node[draw=black, fill=gray!20, rounded corners=2pt,
				minimum width=1.6cm, minimum height=0.4cm, font=\tiny] at (-1, 3.8) {$W_Q,W_K,W_V$};
				\node[font=\tiny, gray] at (-1, 3.45) {frozen};
				
				\node[draw=black, fill=gray!10, rounded corners=2pt,
				minimum width=1.6cm, minimum height=0.4cm, font=\tiny] at (1, 3.8) {$W + AB$};
				\node[font=\tiny, gray] at (1, 3.45) {$r \ll d$};
				
				\draw[arrow] (emb1) -- (trans1);
				
				\node[box] (repr1) at (0, 2.7) {$[h_1, h_2, \ldots, h_T]$};
				\draw[arrow] (trans1) -- (repr1);
				
				\node[box] (out1) at (0, 1.8) {$\hat{y}_t \in \{0,1\}$};
				\draw[arrow] (repr1) -- (out1);
				
				\node[draw=black, fill=gray!10, rounded corners=3pt, line width=0.8pt,
				minimum width=4.2cm, minimum height=0.55cm, font=\tiny] (loss1) at (0, 0.5) {
					$\mathcal{L}_1 = \mathcal{L}_{WCE} + \frac{\mu}{2}\|w - w^{(t)}\|^2$
				};
				\draw[arrow] (out1) -- (loss1);
			\end{scope}
			
			%==============================================================================
			% CLIENT 2
			%==============================================================================
			\begin{scope}[shift={(0, 0)}]
				\node[draw=black, fill=white, rounded corners=5pt, line width=1pt,
				minimum width=4.6cm, minimum height=7.2cm] (c2box) at (0, 3.8) {};
				\node[clientlabel] at (0, 7.2) {Client 2};
				\node[font=\scriptsize, gray] at (0, 6.8) {$\mathcal{D}_2$, $n_2$ samples};
				
				\node[box] (in2) at (0, 6.2) {$[k_1, k_2, \ldots, k_T]$};
				\node[box] (emb2) at (0, 5.4) {Embedding: $k_i \rightarrow \mathbf{e}_i$};
				\draw[arrow] (in2) -- (emb2);
				
				\node[draw=black, fill=gray!5, rounded corners=4pt, line width=0.8pt,
				minimum width=4.3cm, minimum height=1.6cm] (trans2) at (0, 4) {};
				\node[font=\scriptsize\bfseries] at (0, 4.5) {Transformer + LoRA};
				\node[draw=black, fill=gray!20, rounded corners=2pt,
				minimum width=1.6cm, minimum height=0.4cm, font=\tiny] at (-1, 3.8) {$W_Q,W_K,W_V$};
				\node[font=\tiny, gray] at (-1, 3.45) {frozen};
				\node[draw=black, fill=gray!10, rounded corners=2pt,
				minimum width=1.6cm, minimum height=0.4cm, font=\tiny] at (1, 3.8) {$W + AB$};
				\node[font=\tiny, gray] at (1, 3.45) {$r \ll d$};
				\draw[arrow] (emb2) -- (trans2);
				
				\node[box] (repr2) at (0, 2.7) {$[h_1, h_2, \ldots, h_T]$};
				\draw[arrow] (trans2) -- (repr2);
				
				\node[box] (out2) at (0, 1.8) {$\hat{y}_t \in \{0,1\}$};
				\draw[arrow] (repr2) -- (out2);
				
				\node[draw=black, fill=gray!10, rounded corners=3pt, line width=0.8pt,
				minimum width=4.2cm, minimum height=0.55cm, font=\tiny] (loss2) at (0, 0.5) {
					$\mathcal{L}_2 = \mathcal{L}_{WCE} + \frac{\mu}{2}\|w - w^{(t)}\|^2$
				};
				\draw[arrow] (out2) -- (loss2);
			\end{scope}
			
			%==============================================================================
			% DOTS BETWEEN CLIENTS
			%==============================================================================
			\node[font=\Large] at (2.6, 4) {$\cdots$};
			
			%==============================================================================
			% CLIENT K
			%==============================================================================
			\begin{scope}[shift={(5.2, 0)}]
				\node[draw=black, fill=white, rounded corners=5pt, line width=1pt,
				minimum width=4.6cm, minimum height=7.2cm] (c3box) at (0, 3.8) {};
				\node[clientlabel] at (0, 7.2) {Client $K$};
				\node[font=\scriptsize, gray] at (0, 6.8) {$\mathcal{D}_K$, $n_K$ samples};
				
				\node[box] (in3) at (0, 6.2) {$[k_1, k_2, \ldots, k_T]$};
				\node[box] (emb3) at (0, 5.4) {Embedding: $k_i \rightarrow \mathbf{e}_i$};
				\draw[arrow] (in3) -- (emb3);
				
				\node[draw=black, fill=gray!5, rounded corners=4pt, line width=0.8pt,
				minimum width=4.3cm, minimum height=1.6cm] (trans3) at (0, 4) {};
				\node[font=\scriptsize\bfseries] at (0, 4.5) {Transformer + LoRA};
				\node[draw=black, fill=gray!20, rounded corners=2pt,
				minimum width=1.6cm, minimum height=0.4cm, font=\tiny] at (-1, 3.8) {$W_Q,W_K,W_V$};
				\node[font=\tiny, gray] at (-1, 3.45) {frozen};
				\node[draw=black, fill=gray!10, rounded corners=2pt,
				minimum width=1.6cm, minimum height=0.4cm, font=\tiny] at (1, 3.8) {$W + AB$};
				\node[font=\tiny, gray] at (1, 3.45) {$r \ll d$};
				\draw[arrow] (emb3) -- (trans3);
				
				\node[box] (repr3) at (0, 2.7) {$[h_1, h_2, \ldots, h_T]$};
				\draw[arrow] (trans3) -- (repr3);
				
				\node[box] (out3) at (0, 1.8) {$\hat{y}_t \in \{0,1\}$};
				\draw[arrow] (repr3) -- (out3);
				
				\node[draw=black, fill=gray!10, rounded corners=3pt, line width=0.8pt,
				minimum width=4.2cm, minimum height=0.55cm, font=\tiny] (loss3) at (0, 0.5) {
					$\mathcal{L}_K = \mathcal{L}_{WCE} + \frac{\mu}{2}\|w - w^{(t)}\|^2$
				};
				\draw[arrow] (out3) -- (loss3);
			\end{scope}
			
			%==============================================================================
			% DP CLIPPING
			%==============================================================================
			\node[dpbox] (clip1) at (-5.2, -0.5) {Clip: $\|\Delta_1\|_2 \leq C$};
			\node[dpbox] (clip2) at (0, -0.5) {Clip: $\|\Delta_2\|_2 \leq C$};
			\node[dpbox] (clip3) at (5.2, -0.5) {Clip: $\|\Delta_K\|_2 \leq C$};
			
			\draw[bigarrow] (-5.2, 0.1) -- (clip1);
			\draw[bigarrow] (0, 0.1) -- (clip2);
			\draw[bigarrow] (5.2, 0.1) -- (clip3);
			
			%==============================================================================
			% AGGREGATION
			%==============================================================================
			\draw[bigarrow] (clip1.south) -- ++(0, -0.25) -| (0, -1.5);
			\draw[bigarrow] (clip2.south) -- (0, -1.5);
			\draw[bigarrow] (clip3.south) -- ++(0, -0.25) -| (0, -1.5);
			
			\node[draw=black, fill=gray!10, rounded corners=4pt, line width=1pt,
			minimum width=6cm, minimum height=0.7cm, font=\footnotesize] (agg) at (0, -2) {
				$\bar{w} = \sum_{k=1}^{K} \frac{n_k}{n} w_k$
			};
			
			%==============================================================================
			% DP NOISE
			%==============================================================================
			\node[draw=black, fill=gray!15, rounded corners=4pt, line width=1pt,
			minimum width=6cm, minimum height=0.7cm, font=\footnotesize] (noise) at (0, -3) {
				$w^{(t+1)} = \bar{w} + \mathcal{N}(0, \sigma^2 C^2 \mathbf{I})$
			};
			\draw[bigarrow] (agg) -- (noise);
			
			%==============================================================================
			% SERVER - BOTTOM
			%==============================================================================
			\node[serverbox] (server2) at (0, -4) {Server: Global Model $w^{(t+1)}$};
			\draw[bigarrow] (noise) -- (server2);
			
			%==============================================================================
			% LOOP ARROW
			%==============================================================================
			\draw[bigarrow, dashed, rounded corners=6pt]
			(server2.east) -- ++(1, 0) -- ++(0, 13) -- ++(-1, 0) -- (server.east);
			\node[font=\tiny, fill=white, rotate=90] at (8.7, 2.5) {$t \leftarrow t+1$};
			
		\end{tikzpicture}
	}%
	\caption{FlogTinyLLM architecture. Each client processes log keys $[k_1, \ldots, k_T]$ through a transformer with LoRA adapters ($W + AB$, $r \ll d$). FedProx loss includes proximal term $\frac{\mu}{2}\|w - w_{t}\|^2$. Updates are clipped and aggregated with DP noise.}
	\label{fig:architecture}
\end{figure}
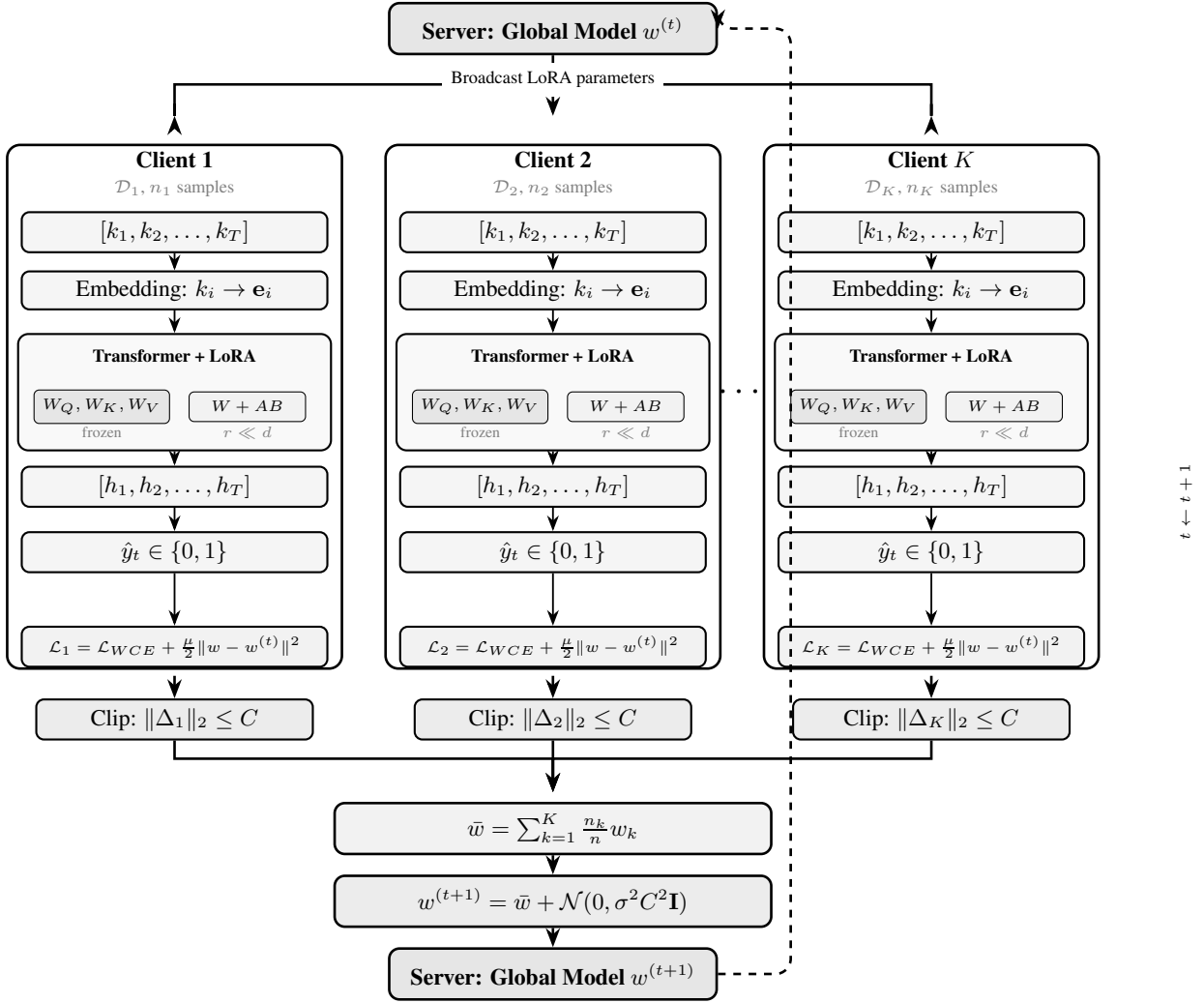

\section{Experimental Setup}

\subsection{Dataset Description}\label{Datasets}

% \subsection{Dataset} \label{Datasets}
The proposed FlogTinyLLM framework is evaluated on two publicly available supercomputer log datasets, as described below

\subsubsection{Thunderbird dataset} \label{T_Datasets}
The Thunderbird dataset is generated on a 4,096-node Dell high-performance computer cluster called Thunderbird located at Sandia National Laboratories (SNL) in Albuquerque \cite{singer2005thunderbird}. It provides about 8000 processors of compute capacity. The computer's aggregated capacity is about 24 terabytes of memory and 60 tera-OPS (trillion operations per second) \cite{singer_corwell2006thunderbird}. The Thunderbird dataset comprises alert and non-alert messages, each tagged with a specific alert category. In the first column of the log, a hyphen ("-") indicates non-alert messages, while other entries represent alert messages. The dataset contains about 211, 212,192 log entries.

\subsubsection{Blue Gene/L(BGL) dataset} \label{B_Datasets}
The BGL log dataset contains system logs from the Blue Gene/L supercomputer at Lawrence Livermore National Laboratory (LLNL), California. Each compute node has a single ASIC with two 700 MHz CPU cores, two floating-point units, 4 MB of embedded cache, and 256 MB of external memory. Nodes connect via high-speed ports in a $32 \times 32 \times 64$ three-dimensional torus network. The system achieves a peak performance of 360 Teraflops and a total memory of 16 Terabytes \cite{llnl_bluegene_l}. The logs are semi-structured, automatically generated messages that contain both alert and non-alert entries. A dash in the first log column indicates a non-alert message. Each message includes fields such as label, epoch timestamp, date, node location, full timestamp, source node, system, component, severity and message body. The dataset contains about 4.7 million log entries and 3,853 log templates. Of these, 348,460 (7.3\%) are labeled as anomalies, and 4,399,503 (92.7\%) as normal \cite{tian2023cldtlog}.

\begin{algorithm2e}[h]
	\caption{FlogTinyLLM: Federated Log Anomaly Detection with LoRA and DP}
	\label{alg:flogtinyllm}
	\DontPrintSemicolon
	\SetAlgoLined
	\SetKwInOut{Input}{Input}
	\SetKwInOut{Output}{Output}
	\SetKwComment{tcp}{$\triangleright$~}{}
	
	\Input{$C$ clients, $T$ rounds, $E$ local epochs, learning rate $\eta$, FedProx coefficient $\mu$, clipping bound $C$, noise multiplier $\sigma$, LoRA rank $r \ll d$}
	\Output{Trained global model $w^{(T)}$}
	
	\BlankLine
	\tcp{Phase 1: Initialization}
	$w^{(0)} \leftarrow$ pre-trained transformer\;
	Attach LoRA adapters: $A \in \mathbb{R}^{d \times r}$, $B \in \mathbb{R}^{r \times d}$\;
	Freeze attention weights: $W_Q$, $W_K$, $W_V$\;
	
	\BlankLine
	\For{$t = 0$ \KwTo $T-1$}{
		
		\tcp{Phase 2: Broadcast}
		Server sends $w^{(t)}$ to all $C$ clients\;
		
		\BlankLine
		\tcp{Phase 3: Local Training}
		\For{each client $C = 1, \ldots, C$ \textbf{in parallel}}{
			
			$w_k \leftarrow w^{(t)}$\;
			
			\For{epoch $e = 1$ \KwTo $E$}{
				\For{batch $\big([k_1, k_2, \ldots, k_T],\, y\big) \in \mathcal{D}_k$}{
					
					\tcp{Forward pass}
					$\mathbf{e}_i \leftarrow \textsc{Embed}(k_i)$, \quad $i = 1, \ldots, T$\;
					$[h_1, \ldots, h_T] \leftarrow \textsc{Transformer}\big(\mathbf{e}_1, \ldots, \mathbf{e}_T;\, W + AB\big)$\;
					$\hat{y} \leftarrow \textsc{Softmax}\big(\textsc{Linear}(h_T)\big)$\;
					
					\tcp{FedProx loss}
					$\mathcal{L}_{CE} \leftarrow -y \log \hat{y} - (1-y) \log(1-\hat{y})$\;
					$\mathcal{L}_{prox} \leftarrow \frac{\mu}{2} \|w_k - w^{(t)}\|_2^2$\;
					$\mathcal{L}_k \leftarrow \mathcal{L}_{CE} + \mathcal{L}_{prox}$\;
					
					\tcp{Gradient update}
					$w_k \leftarrow w_k - \eta \, \nabla_{w_k} \mathcal{L}_k$\;
				}
			}
			
			\tcp{Compute and clip update}
			$\Delta_k \leftarrow w_k - w^{(t)}$\;
			$\Delta_k \leftarrow \Delta_k \cdot \min\Big(1,\, \frac{C}{\|\Delta_k\|_2}\Big)$\;
			
			Send $\Delta_k$ and $n_k = |\mathcal{D}_k|$ to server\;
		}
		
		\BlankLine
		\tcp{Phase 4: Aggregation with Differential Privacy}
		$n \leftarrow \sum_{k=1}^{K} n_k$\;
		$\bar{w} \leftarrow w^{(t)} + \sum_{k=1}^{K} \frac{n_k}{n} \, \Delta_k$\;
		$w^{(t+1)} \leftarrow \bar{w} + \mathcal{N}\big(0,\, \sigma^2 C^2 \mathbf{I}\big)$\;
	}
	
	\BlankLine
	\Return $w^{(T)}$\;
\end{algorithm2e}

\subsection{Data Processing}

After log parsing, the structured log entries are segmented using a sliding-window technique to form sequences of log keys. For the Thunderbird dataset, a five-minute sliding window with a one-minute step size is applied, and for the BGL dataset, an eight-minute window with a two-minute step size is used. This segmentation captures system activity within each window, producing sequences of log keys. The sequences are then distributed using the Round-Robin algorithm to  14 clients for the Thunderbird dataset  and 15 clients for the BGL dataset forming individual client datasets $\{D_i\}_{i=1}^{N}$, each consisting of log key sequences $\{k_{i1}, k_{i2}, \dots, k_{in}\},\quad i = 1,2,\dots,n$.  This enables FlogTinyLLM to detect diverse anomaly patterns throughout clients using the tiny language models enhanced with LoRA.

\subsection{Model Specifications}

% oooooooooooooooo
\begin{table*}[h]
	\centering
	\caption{Experimental Configuration for FlogTinyLLM on The Thunderbird Dataset}
	\label{tab:experiment_config}
	\renewcommand{\arraystretch}{1.2}
	\begin{tabular}{lcccccc}
		\toprule
		\textbf{Category} & \textbf{Parameter} & \textbf{Value} \\
		\midrule
		\multirow{2}{*}{\textbf{Dataset}} 
		& Maximum Samples & 4,500,000 \\
		& Minimum Anomaly Rate per Node & 0.001 \\
		\midrule
		\multirow{3}{*}{\textbf{LoRA}} 
		& Rank ($r$) & 8 \\
		& Scaling Factor ($\alpha$) & 32 \\
		& Dropout Rate & 0.1 \\
		\midrule
		\multirow{4}{*}{\textbf{Federated Learning}} 
		& Number of Clients & 14 \\
		& Communication Rounds ($T$) & 10 \\
		& Participation Rate ($q$) & 50\% \\
		& Proximal Coefficient ($\mu$) & 0.01 \\
		\midrule
		\multirow{3}{*}{\textbf{Sliding Window}} 
		& Window Size & 5 min \\
		& Step Size & 1 min \\
		& Minimum Logs per Window & 5 \\
		\midrule
		\multirow{7}{*}{\textbf{Training}} 
		& Batch Size & 8 \\
		& Learning Rate & $2 \times 10^{-5}$ \\
		& Weight Decay & 0.01 \\
		& Local Epochs ($E$) & 10 \\
		& Warmup Ratio & 0.1 \\
		& Gradient Accumulation Steps & 2 \\
		& Maximum Gradient Norm & 1.0 \\
		\midrule
		\multirow{5}{*}{\textbf{Differential Privacy}} 
		& Privacy Budget ($\varepsilon$) & 10.0 \\
		& Delta ($\delta$) & $10^{-5}$ \\
		& Clipping Bound ($C$) & 1.0 \\
		& Noise Multiplier ($\sigma$) & 0.01 \\
		
		\bottomrule
	\end{tabular}
\end{table*}

Table \ref{tab:experiment_config} outlines the experimental setup for FlogTinyLLM on the Thunderbird dataset, with parameters grouped as follows: The training data contains up to 4,500,000 samples and a minimum anomaly rate of 0.001 per node. The LoRA parameters include rank 8, a scaling factor of 32, and a dropout rate of 0.1. Federated learning settings are 14 clients, 10 communication rounds, 50\% participation rate, and a proximal coefficient of 0.01. Sliding Window parameters are a 5-minute window size, a 1-minute step size, and at least 5 logs per window. Training parameters are batch size 8, learning rate $2 \times 10^{-5}$, weight decay 0.01, 10 local epochs, warmup ratio 0.1, 2 gradient accumulation steps, and maximum gradient norm 1.0. The Differential Privacy parameters are a privacy budget of 10.0, a delta of $10^{-5}$, a clipping bound of 1.0, and a noise multiplier of 0.01.

% 000000000000000000000000000000000000000000
\begin{table*}[h]
	\centering
	\caption{Experimental Configuration for FlogTinyLLM on The BGL Dataset}
	\label{tab:fl_architecture}
	\vspace{4pt}
	\small
	\begin{tabular}{lll}
		\toprule
		\textbf{Category} & \textbf{Parameter} & \textbf{Value} \\
		\midrule
		
		\multirow{1}{*}{\textbf{Dataset}}
		& Maximum Samples     & 500{,}000 \\
		\midrule
		
		\multirow{4}{*}{\textbf{LoRA Adaptation}}
		& Rank ($r$)          & 8 \\
		& Scaling Factor ($\alpha$) & 32 \\
		& Dropout             & 0.1 \\
		\midrule
		
		\multirow{4}{*}{\textbf{Federated Learning}}
		& Number of Clients   & 15 \\
		& Communication Rounds & 20 \\
		& Participation Rate  & 0.7 (70\%) \\
		& FedProx Proximal $\mu$ & 0.001 \\
		\midrule
		
		\multirow{3}{*}{\textbf{Sliding Window}}
		& Window Size         & 8 minutes \\
		& Step Size           & 2 minutes \\
		& Minimum Logs per Window & 6 \\
		\midrule
		
		\multirow{7}{*}{\textbf{Local Training}}
		& Batch Size          & 8 \\
		& Learning Rate       & $3 \times 10^{-5}$ \\
		& Local Epochs        & 5 \\
		& Weight Decay        & 0.01 \\
		& Warmup Ratio        & 0.1 \\
		& Max Gradient Norm   & 1.0 \\
		& Gradient Accumulation Steps & 1 \\
		\midrule
		
		\multirow{4}{*}{\textbf{Differential Privacy}}
		& Privacy Budget ($\varepsilon$) & 10.0 \\
		& Delta ($\delta$)    & $1 \times 10^{-5}$ \\
		& Clipping Bound ($C$) & 1.0 \\
		& Noise Multiplier ($\sigma$) & 1.5 \\
		
		\bottomrule
	\end{tabular}
	\vspace{6pt}
	
%	\raggedright
	
\end{table*}

\begin{table*}[h]
	\centering
	\caption{FlogTinyLLM architecture and parameter efficiency comparison on the Thunderbird dataset.}
	\label{tab:architecture}
	\begin{threeparttable}
		\begin{tabular}{@{}lccccr@{}}
			\toprule
			\textbf{Model} & \textbf{Base Params} & \textbf{Trainable} & \textbf{Ratio (\%)} & \textbf{Memory} & \textbf{$\|$noise$\|_2$} \\
			\midrule
			% \rowcolor{lightblue}
			Microsoft Phi-1.5 & 1.3B & 3,149,824 & 0.44 & 2,907.6 MB & 5.071 \\
			% \rowcolor{lightorange}
			DeepSeek-R1-Qwen & 1.5B & 1,494,016 & 0.17 & 2,978.2 MB & 3.492 \\
			% \rowcolor{lightgreen}
			Facebook OPT-1.3B & 1.3B & 2,363,392 & 0.33 & 2,820.2 MB & 4.392 \\
			% \rowcolor{lightpink}
			TinyLlama-1.1B & 1.1B & 1,536,000 & 0.28 & 4,221.3 MB & 3.540 \\
			\bottomrule
		\end{tabular}
		\begin{tablenotes}
			\small
			\item Note: All models use LoRA with $r=8$, $\alpha=32$. Noise std = 0.002857 for all models.
		\end{tablenotes}
	\end{threeparttable}
\end{table*}

Table \ref{tab:experiment_config} and Table~\ref{tab:fl_architecture}  summarize the experimental setup for the FlogTinyLLM framework on both datasets. Table \ref{tab:architecture} presents architectural and parameter efficiency across the four FlogTinyLLM models on the Thunderbird dataset. Reported metrics are: base parameters (billions), trainable parameters, trainable parameter ratio (\%), memory Usage (MB), and noise. All models employ LoRA (r - 8, $\alpha$ - 32, noise std - 0.002857). These metrics directly quantify the trade-offs between memory efficiency and noise. Phi-1.5 has a base parameter of 1.3B but is reduced to a trainable parameter of 3,149,824 due to LoRA adoption, representing 0.44\% of the total base parameter. Phi-1.5 used a memory of 2,907.6 MB, representing the second least for the FlogTinyLLM training, which indicates moderate resource requirements for execution. DeepSeek-R1 has a base parameter count of 1.5B and a trainable parameter count of 1,494,016 (second-least), representing a ratio of 0.17\% (least) and 2,978.2 MB of memory utilization (third-least). OPT-1.3B has a base parameter of 1.3B with a trainable parameter of 2,363,392 (third most), representing a ratio of 0.33\% in the total base parameter, with the least memory utilization of 2,820.2 MB. TinyLlama-1.1B has a base parameter count of 1.1B and a trainable parameter count of 1,536,000 (minimum), representing a ratio of 0.28\% (second-lowest), with the highest memory utilization of 4,221.3 MB. FlogTinyLLM has base parameter range 1.1B-1.5B; Trainable 1,494,016-3,149,824 (ratios 0.17\%-0.44\%); Memory 2,820.2-4,221.3 MB; $||\text{noise}||_2$ 3.492-5.071. Specifically, Phi-1.5 has the highest trainable count/ratio/noise but moderate memory. In contrast, DeepSeek-R1 has the highest base but the lowest ratio/noise, making it feasible for multi-GPU setups that demand minimal noise. Lastly, TinyLlama, with the lowest base/trainable but highest memory.

\subsection{Training}

Algorithm~\ref{alg:flogtinyllm} presents the step-by-step training procedure of FlogTinyLLM. The process runs in four phases that repeat over $T$ communication rounds between a central server and $C$ participating clients.

\paragraph{Phase 1: Initialization.} Before any communication begins, the server loads a pretrained language model and attaches LoRA adapter matrices $A \in \mathbb{R}^{d \times r}$ and $B \in \mathbb{R}^{r \times d}$ to each attention layer. The original attention projection weights $W_Q$, $W_K$, and $W_V$ of the language models are then frozen. From this point onward, training modifies only  matrices $A$ and $B$ \cite{sanowl_lora}.

\paragraph{Phase 2: Broadcast.} At the start of each round $t$, the server sends the current global
parameters $W_{t}$ to every participating client. Each client trains its local model with these received
parameters\cite{singamsetty2025federated_learning}, so that all participating clients begin the round from the same starting point.

\paragraph{Phase 3: Local Training.} All $C$ participating clients train in parallel. For a single batch containing a sequence $[k_1, k_2, \ldots, k_T]$ and its ground-truth label $y$, the forward pass works as follows. Each
log key $k_i$ is mapped to a dense embedding $\mathbf{e}_i$. The sequence of embeddings passes
through the transformer, where the frozen weights $W$ are combined with the trainable LoRA product
$AB$ to produce contextual representations $[h_1, \ldots, h_T]$. The final representation
$h_T$ is fed through a linear layer and softmax to yield the prediction $\hat{y}$. The client then computes its local loss. This loss has two parts. The first is a weighted  cross-entropy term

\begin{equation}
	\mathcal{L}_{\text{WCE}} = -\frac{1}{n_k} \sum_{j=1}^{n_k} 
	\Big[ w_1 \cdot y_j \log \hat{y}_j 
	+ w_0 \cdot (1 - y_j) \log(1 - \hat{y}_j) \Big]
	\label{eq:wce}
\end{equation} 

\noindent 
where $y_j \in \{0, 1\}$ is the ground-truth label for sequence $j$, $\hat{y}_j$ is the model's 
predicted probability that sequence $j$ is anomalous, $w_1$ is the weight assigned to the anomalous class, 
and $w_0$ is the weight assigned to the normal class. The class weights are typically set inversely proportional to the class frequency in the training  data~\cite{king2001logistic}. This measures how accurately the model classifies the sequence. The second is the FedProx proximal term $\frac{\mu}{2} |W - W_t|^2$ that penalizes the local parameters $w_k$ for drifting away from the global parameters $W_{t}$ \cite{keerthika2025proximal_fl}. The total loss is their sum. The client updates its parameters by taking a gradient step: $w_k \leftarrow w_k - \eta \,\nabla_{w_k} \mathcal{L}_k$. After all local epochs are complete, the client computes its total update $\Delta_k = w_k - w_{t}$, which represents how far its parameters have moved from the global model it started with. The update is then clipped \cite{li2022preserving_fl_privacy}:

$\Delta_k \leftarrow \Delta_k \cdot \min\big(1, \frac{C}{|\Delta_k|_2}\big)$.
If the update is already within the bound, it passes through unchanged. If it exceeds the bound, it is
scaled down proportionally. The clipped updates $\Delta_k$  are sent to the server.

\paragraph{Phase 4: Aggregation with Differential Privacy.} The server collects all clipped updates
and combines them into a single global update. It first computes the total number of samples across
all clients: $|D| = \sum_{k=1}^{K} n_k$. The aggregated model is then formed by adding a weighted
average of the client updates to the current global model:
$\bar{w} = w_{t} + \sum_{k=1}^{K} \frac{|D_i|}{|D|} . \Delta_k$. The contribution of each client is proportional to the amount of data it holds, so clients with larger datasets have a greater influence on the result \cite{zhang2025secure_aggregation_fl}. To make the aggregated model the next global model, the server adds gaussian noise: $w_{t+1} = \bar{w} + \mathcal{N}(0, \sigma^2 C^2 \mathbf{I})$. The noise magnitude is determined by two quantities: the noise multiplier $\sigma$, which controls the overall noise level, and the clipping bound $C$, which limits the client updates. The product $\sigma^2 C^2$ calibrates the noise so that it is large enough to mask the contribution of any individual client, satisfying the differential privacy guarantee \cite{paverd2015enhancing}. The server then stores $w_{t+1}$ as the new global model, and the next round begins from Phase 2. After $T$ rounds, the server returns the final model $W_{T}$, which has absorbed patterns from all client log data.

\section{Experimental Results}

This section outlines the experimental results of FlogTinyLLM  on the Thunderbird and the BGL datasets
\subsection{Performance Analysis of FlogTinyLLM on the Thunderbird Dataset}

Table \ref{tab:comprehensive_results} oulines the comprehensive perfomance of FlogTinyLLM on the Thunderbird dataset

\begin{table*}[h]
	\centering
	\caption{Comprehensive Performance Metrics: Final Values and Best Achieved}
	\label{tab:comprehensive_results}
	\footnotesize
	\begin{tabular}{@{}llcccccc@{}}
		\toprule
		\textbf{Model} & \textbf{Metric} & \textbf{Accuracy} & \textbf{Precision} & \textbf{Recall} & \textbf{F1} & \textbf{ROC-AUC} \\
		                         & \textbf{Type}    &                               &                                &                           & \textbf{Score} &  \\
		\midrule
		\multirow{3}{*}{\textbf{FlogTinyLLM: Phi-1.5}} 
		& Final Value & 0.9190 & 0.9997 & 0.9182 & 0.9572 & 0.9963 \\
		& Best Value & 0.9986 & 0.9998 & 0.9997 & 0.9993 & 0.9963 \\
		& Best Round & 1 & 3 & 1 & 1 & 10 \\
		\midrule
		\multirow{3}{*}{\textbf{FlogTinyLLM: DeepSeek-R1}} 
		& Final Value & 0.9984 & 0.9997 & 0.9987 & 0.9992 & 0.9982 \\
		& Best Value & 0.9985 & 0.9998 & 0.9991 & 0.9992 & 0.9982 \\
		& Best Round & 7 & 9 & 1 & 7 & 10 \\
		\midrule
		\multirow{3}{*}{\textbf{FlogTinyLLM: OPT-1.3B}} 
		& Final Value & 0.9984 & 0.9998 & 0.9986 & 0.9992 & 0.9967 \\
		& Best Value & 0.9988 & 0.9998 & 0.9997 & 0.9994 & 0.9967 \\
		& Best Round & 5 & 10 & 1 & 5 & 10 \\
		\midrule
		\multirow{3}{*}{\textbf{FlogTinyLLM: TinyLlama-1.1B}} 
		& Final Value & 0.9571 & 0.9994 & 0.9571 & 0.9778 & 0.9912 \\
		& Best Value & 0.9583 & 0.9994 & 0.9585 & 0.9784 & 0.9913 \\
		& Best Round & 5 & 10 & 5 & 5 & 9 \\
		\bottomrule
	\end{tabular}
\end{table*}

\subsubsection{Accuracy Convergence Dynamics}

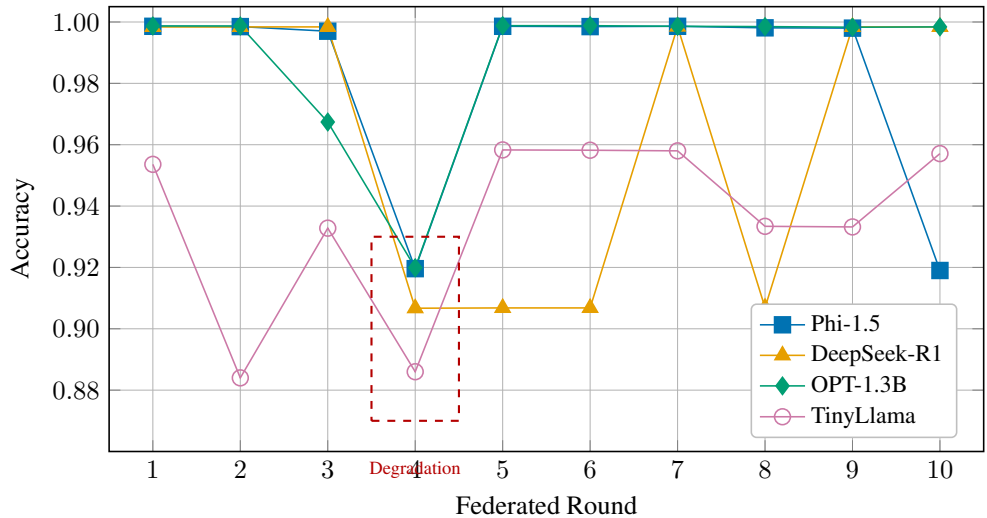
\begin{figure*}[h]
	\centering
	\begin{tikzpicture}
		\begin{axis}[
			width=0.8\textwidth,
			height=7.5cm,
			xlabel={Federated Round},
			ylabel={Accuracy},
			xmin=0.5, xmax=10.5,
			ymin=0.86, ymax=1.005,
			xtick={1,2,3,4,5,6,7,8,9,10},
			ytick={0.88,0.90,0.92,0.94,0.96,0.98,1.00},
			yticklabel style={/pgf/number format/.cd, fixed, fixed zerofill, precision=2},
			legend pos=south east,
			legend style={font=\small, fill=white, fill opacity=0.9, draw=gray!50, rounded corners=2pt},
			legend cell align={left},
			grid=both,
			grid style={line width=0.1pt, draw=gridgray},
			major grid style={line width=0.2pt, draw=gray!60},
			mark size=2.5pt,
			line width=0.6pt,
			every axis plot/.append style={mark options={solid}},
			clip=false
			]
			
			% Phi-1.5 Accuracy - VERIFIED DATA
			\addplot[color=phi, mark=square*, mark options={fill=phi, scale=1.2}] coordinates {
				(1,0.9986) (2,0.9985) (3,0.9970) (4,0.9196) (5,0.9986) 
				(6,0.9985) (7,0.9986) (8,0.9981) (9,0.9980) (10,0.9190)
			};
			\addlegendentry{Phi-1.5}
			
			% DeepSeek-R1 Accuracy - VERIFIED DATA
			\addplot[color=deepseek, mark=triangle*, mark options={fill=deepseek, scale=1.3}] coordinates {
				(1,0.9984) (2,0.9984) (3,0.9984) (4,0.9067) (5,0.9068) 
				(6,0.9068) (7,0.9985) (8,0.9069) (9,0.9984) (10,0.9984)
			};
			\addlegendentry{DeepSeek-R1}
			
			% OPT-1.3B Accuracy - VERIFIED DATA
			\addplot[color=opt, mark=diamond*, mark options={fill=opt, scale=1.3}] coordinates {
				(1,0.9987) (2,0.9987) (3,0.9674) (4,0.9199) (5,0.9988) 
				(6,0.9988) (7,0.9986) (8,0.9985) (9,0.9983) (10,0.9984)
			};
			\addlegendentry{OPT-1.3B}
			
			% TinyLlama Accuracy - VERIFIED DATA
			\addplot[color=tinyllama, mark=o, mark options={fill=tinyllama, scale=1.2}] coordinates {
				(1,0.9536) (2,0.8840) (3,0.9328) (4,0.8860) (5,0.9583) 
				(6,0.9582) (7,0.9580) (8,0.9334) (9,0.9332) (10,0.9571)
			};
			\addlegendentry{TinyLlama}
			
			% Annotation for performance drops
			\draw[darkred, thick, dashed] (axis cs:3.5,0.87) rectangle (axis cs:4.5,0.93);
			\node[darkred, font=\scriptsize, anchor=north] at (axis cs:4,0.86) {Degradation};
			
		\end{axis}
	\end{tikzpicture}
	\caption{Accuracy convergence across federated rounds. Notable performance degradation occurs at Round 4 for multiple models due to client heterogeneity effects.}
	\label{fig:accuracy_convergence}
\end{figure*}

Figure \ref{fig:accuracy_convergence} illustrates the evolution of the accuracy of the FlogTinyLLM variants across ten federated learning rounds on the Thunderbird dataset. The x-axis represents the federated rounds, ranging from 1 to 10, while the y-axis denotes accuracy, scaled from 0.86 to 1.00 with increments that emphasize performance variations. Each model is depicted with a unique marker and color for clarity: Phi-1.5 (square markers), DeepSeek-R1 (triangle markers), OPT-1.3B (diamond markers), and TinyLlama (circle markers). The plot incorporates a grid for precise visual alignment and highlights a region of notable degradation around federated round 4, attributed to client heterogeneity effects, as indicated by a dashed rectangle and annotation.

The Phi-1.5 model exhibits high initial accuracy, commencing at 0.9986 in Round 1 and maintaining near-perfect performance through Round 2 (0.9985). A minor dip occurs in Round 3 (0.9970), followed by a substantial decline to 0.9196 in Round 4, which represents a relative drop from the preceding round. This perturbation is transient, with recovery evident in Round 5 (0.9986), after which it stabilizes around 0.9985-0.9986 until Round 7. 

In contrast, the DeepSeek-R1 model demonstrates consistent high accuracy in the early rounds, holding steady at 0.9984 from Rounds 1 through 3. A sharp decline in Round 4 (0.9067). Recovery is observed in Round 7 (0.9985), albeit with subsequent oscillations: a return to the depressed level in Round 8 (0.9069), resurgence to 0.9984 in Round 9, and stabilization at 0.9984 in Round 10.

The OPT-1.3B model starts with exemplary performance at 0.9987 in Rounds 1 and 2, then experiences a moderate decline to 0.9674 in Round 3 and further to 0.9199 in Round 4. Post-Round 4, the model rebounds to 0.9988 in Rounds 5 and 6, maintaining accuracy with minimal variance: 0.9986 (Round 7), 0.9985 (Round 8), 0.9983 (Round 9), and 0.9984 (Round 10). This trajectory exemplifies rapid convergence and sustained stability after initial fluctuations, highlighting superior adaptability to federated heterogeneity compared to its counterparts. In contrast, TinyLlama, which commenced at 0.9536 in Round 1, faced a more volatile trajectory, ultimately recovering at a rate approximately half that of OPT-1.3B. Specifically, OPT-1.3B recovered twice as fast as TinyLlama from its accuracy dip in Round 4, demonstrating its robustness and efficiency in adapting to federated learning challenges.

TinyLlama, conversely, displays greater volatility throughout the training process, starting at 0.9536 in Round 1 and dropping to 0.8840 in Round 2. It partially recovers to 0.9328 in Round 3, only to degrade again to 0.8860 in Round 4. Subsequent rounds show incremental improvement, peaking at 0.9583 in Round 5, with minor decrements to 0.9582 (Round 6), 0.9580 (Round 7), followed by dips to 0.9334 (Round 8) and 0.9332 (Round 9), before a slight uptick to 0.9571 in Round 10. Unlike the other models, TinyLlama's accuracy remains consistently below 0.96, exhibiting oscillatory dynamics without achieving near-unity convergence.

\subsubsection{F1-Score Evolution}

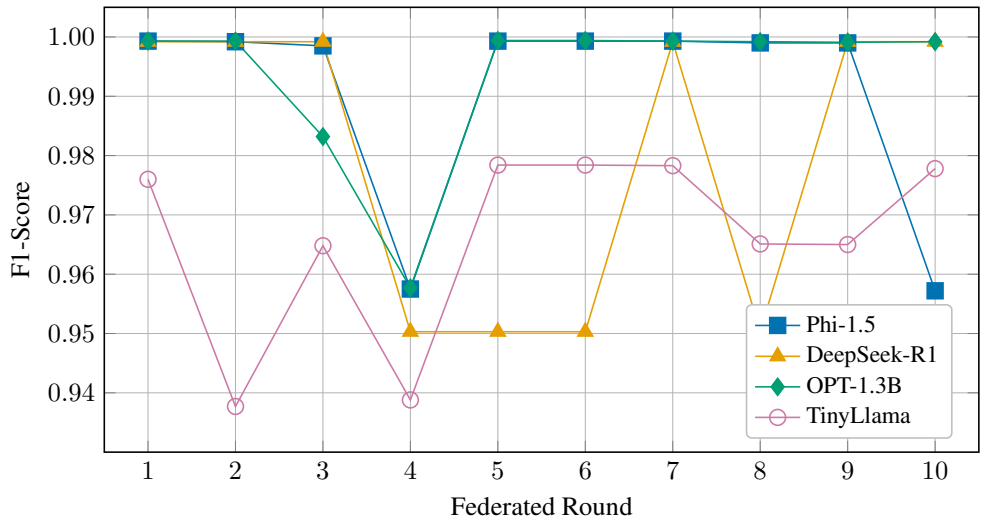
\begin{figure*}[h]
	\centering
	\begin{tikzpicture}
		\begin{axis}[
			width=0.8\textwidth,
			height=7.5cm,
			xlabel={Federated Round},
			ylabel={F1-Score},
			xmin=0.5, xmax=10.5,
			ymin=0.93, ymax=1.005,
			xtick={1,2,3,4,5,6,7,8,9,10},
			ytick={0.94,0.95,0.96,0.97,0.98,0.99,1.00},
			yticklabel style={/pgf/number format/.cd, fixed, fixed zerofill, precision=2},
			legend pos=south east,
			legend style={font=\small, fill=white, fill opacity=0.9, draw=gray!50, rounded corners=2pt},
			legend cell align={left},
			grid=both,
			grid style={line width=0.1pt, draw=gridgray},
			major grid style={line width=0.2pt, draw=gray!60},
			mark size=2.5pt,
			line width=0.6pt
			]
			
			% Phi-1.5 F1 - VERIFIED DATA
			\addplot[color=phi, mark=square*, mark options={fill=phi, scale=1.2}] coordinates {
				(1,0.9993) (2,0.9992) (3,0.9985) (4,0.9575) (5,0.9993) 
				(6,0.9993) (7,0.9993) (8,0.9990) (9,0.9990) (10,0.9572)
			};
			\addlegendentry{Phi-1.5}
			
			% DeepSeek-R1 F1 - VERIFIED DATA
			\addplot[color=deepseek, mark=triangle*, mark options={fill=deepseek, scale=1.3}] coordinates {
				(1,0.9992) (2,0.9992) (3,0.9992) (4,0.9503) (5,0.9503) 
				(6,0.9503) (7,0.9992) (8,0.9504) (9,0.9992) (10,0.9992)
			};
			\addlegendentry{DeepSeek-R1}
			
			% OPT-1.3B F1 - VERIFIED DATA
			\addplot[color=opt, mark=diamond*, mark options={fill=opt, scale=1.3}] coordinates {
				(1,0.9993) (2,0.9993) (3,0.9832) (4,0.9577) (5,0.9994) 
				(6,0.9994) (7,0.9993) (8,0.9992) (9,0.9991) (10,0.9992)
			};
			\addlegendentry{OPT-1.3B}
			
			% TinyLlama F1 - VERIFIED DATA
			\addplot[color=tinyllama, mark=o, mark options={fill=tinyllama, scale=1.2}] coordinates {
				(1,0.9760) (2,0.9377) (3,0.9648) (4,0.9388) (5,0.9784) 
				(6,0.9784) (7,0.9783) (8,0.9651) (9,0.9650) (10,0.9778)
			};
			\addlegendentry{TinyLlama}
			
		\end{axis}
	\end{tikzpicture}
	\caption{F1-Score evolution demonstrating model robustness.}
	\label{fig:f1_convergence}
\end{figure*}

Figure \ref{fig:f1_convergence} compares the F1-score of the four models in the FlogTinyLLM architecture across 10 federated learning rounds.
Compared with the other models, Phi-1.5 maintains high stability in the early rounds and shows a strong recovery after its first performance drop at round 4. However, its notable declines at rounds 4 and 10 distinguish it from models with fewer fluctuations, resulting in the lowest final F1-score among the three top-performing models.
DeepSeek-R1, relative to Phi-1.5,  shows early high performance but experiences a more prolonged dip from rounds 4-6. Its pattern of alternating recoveries and drops makes its stability intermediate between Phi-1.5 and OPT-1.3B.
OPT-1.3B stands out for its superior convergence stability compared to Phi-1.5 and DeepSeek-R1. While all three models recover after mid drops, OPT-1.3B sustains the highest and most stable F1-scores through the final rounds, distinguishing itself as the most reliable among the top models.
Across F1-score evolution, OPT-1.3B and Phi-1.5 demonstrate higher performance and convergence stability, while TinyLlama consistently underperforms with greater score variance. DeepSeek-R1 exhibits oscillating performance, recovering at later rounds.

\subsubsection{Roc-Auc Progression}

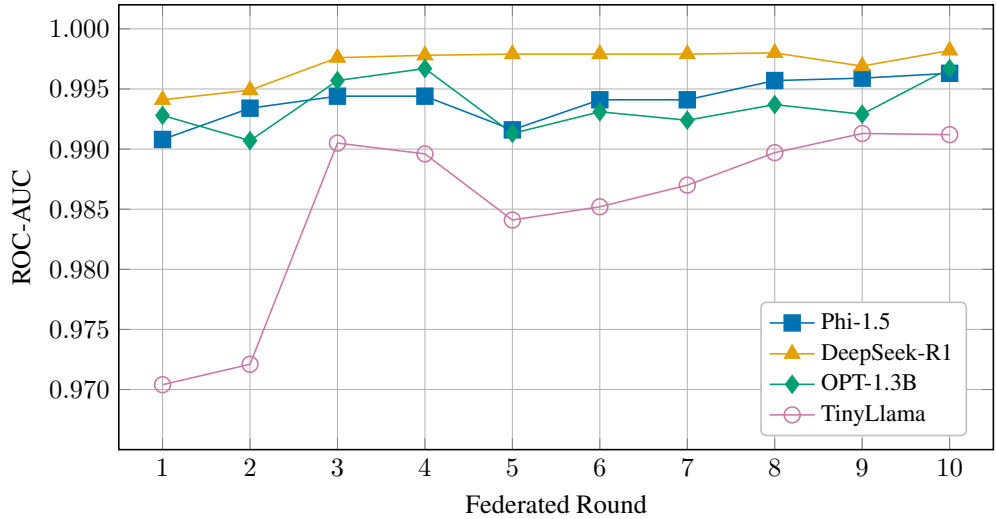
\begin{figure*}[h]
	\centering
	\begin{tikzpicture}
		\begin{axis}[
			width=0.8\textwidth,
			height=7.5cm,
			xlabel={Federated Round},
			ylabel={ROC-AUC},
			xmin=0.5, xmax=10.5,
			ymin=0.965, ymax=1.002,
			xtick={1,2,3,4,5,6,7,8,9,10},
			ytick={0.97,0.975,0.98,0.985,0.99,0.995,1.00},
			yticklabel style={/pgf/number format/.cd, fixed, fixed zerofill, precision=3},
			legend pos=south east,
			legend style={font=\small, fill=white, fill opacity=0.9, draw=gray!50, rounded corners=2pt},
			legend cell align={left},
			grid=both,
			grid style={line width=0.1pt, draw=gridgray},
			major grid style={line width=0.2pt, draw=gray!60},
			mark size=2.5pt,
			line width=0.6pt
			]
			
			% Phi-1.5 AUC - VERIFIED DATA
			\addplot[color=phi, mark=square*, mark options={fill=phi, scale=1.2}] coordinates {
				(1,0.9908) (2,0.9934) (3,0.9944) (4,0.9944) (5,0.9916) 
				(6,0.9941) (7,0.9941) (8,0.9957) (9,0.9959) (10,0.9963)
			};
			\addlegendentry{Phi-1.5}
			
			% DeepSeek-R1 AUC - VERIFIED DATA
			\addplot[color=deepseek, mark=triangle*, mark options={fill=deepseek, scale=1.3}] coordinates {
				(1,0.9941) (2,0.9949) (3,0.9976) (4,0.9978) (5,0.9979) 
				(6,0.9979) (7,0.9979) (8,0.9980) (9,0.9969) (10,0.9982)
			};
			\addlegendentry{DeepSeek-R1}
			
			% OPT-1.3B AUC - VERIFIED DATA
			\addplot[color=opt, mark=diamond*, mark options={fill=opt, scale=1.3}] coordinates {
				(1,0.9928) (2,0.9907) (3,0.9957) (4,0.9967) (5,0.9913) 
				(6,0.9931) (7,0.9924) (8,0.9937) (9,0.9929) (10,0.9967)
			};
			\addlegendentry{OPT-1.3B}
			
			% TinyLlama AUC - VERIFIED DATA
			\addplot[color=tinyllama, mark=o, mark options={fill=tinyllama, scale=1.2}] coordinates {
				(1,0.9704) (2,0.9721) (3,0.9905) (4,0.9896) (5,0.9841) 
				(6,0.9852) (7,0.9870) (8,0.9897) (9,0.9913) (10,0.9912)
			};
			\addlegendentry{TinyLlama}
			
		\end{axis}
	\end{tikzpicture}
	\caption{ROC-AUC progression }
	\label{fig:auc_convergence}
\end{figure*}

Figure \ref{fig:auc_convergence} shows how discriminative ability changes for the FlogtinyLLM variant on the Thunderbird dataset.
Phi-1.5, though improved, is outperformed by DeepSeek-R1 in all rounds. Its minor gains place it above TinyLlama and OPT-1.3B in the final ROC-AUC, but it remains always below DeepSeek-R1.
DeepSeek-R1 leads its cohort in discriminative performance, maintaining the highest ROC-AUC in every round.
TinyLlama starts with the lowest value (0.9704), but improves the most, reaching 0.9912 by round 10. It gains sharply between rounds 2 and 3 (0.9721 to 0.9905), then stabilizes in the 0.984-0.991 range.

\begin{figure}[H]
	\centering
	\begin{tikzpicture}
		\begin{axis}[
			width=0.95\textwidth,
			height=8cm,
			xlabel={Federated Round ($\varepsilon$ spent)},
			ylabel={F1-Score},
			xmin=0.5, xmax=10.5,
			ymin=0.92, ymax=1.005,
			xtick={1,2,3,4,5,6,7,8,9,10},
			xticklabels={1 ($\varepsilon\!=\!1$),2 ($\varepsilon\!=\!2$),3 ($\varepsilon\!=\!3$),4 ($\varepsilon\!=\!4$),5 ($\varepsilon\!=\!5$),6 ($\varepsilon\!=\!6$),7 ($\varepsilon\!=\!7$),8 ($\varepsilon\!=\!8$),9 ($\varepsilon\!=\!9$),10 ($\varepsilon\!=\!10$)},
			x tick label style={font=\tiny, rotate=45, anchor=east},
			legend style={at={(0.02,0.02)}, anchor=south west, font=\small, draw=black, fill=white, fill opacity=0.9},
			grid=major,
			grid style={dashed, gray!40},
			title={\textbf{F1-Score vs.\ Privacy Budget Consumption on the Thunderbird dataset}},
			mark size=2.5pt,
			line width=1.2pt,
			]
			
			% Phi-1.5
			\addplot[color=phi, mark=square*] coordinates {
				(1,0.9993) (2,0.9992) (3,0.9985) (4,0.9575) (5,0.9993) (6,0.9993) (7,0.9993) (8,0.9990) (9,0.9990) (10,0.9572)
			};
			\addlegendentry{Phi-1.5}
			
			% DeepSeek
			\addplot[color=deepseek, mark=triangle*] coordinates {
				(1,0.9992) (2,0.9992) (3,0.9992) (4,0.9503) (5,0.9503) (6,0.9503) (7,0.9992) (8,0.9504) (9,0.9992) (10,0.9992)
			};
			\addlegendentry{DeepSeek-R1-Distill-Qwen-1.5B}
			
			% OPT-1.3B
			\addplot[color=opt, mark=diamond*] coordinates {
				(1,0.9993) (2,0.9993) (3,0.9832) (4,0.9577) (5,0.9994) (6,0.9994) (7,0.9993) (8,0.9992) (9,0.9991) (10,0.9992)
			};
			\addlegendentry{OPT-1.3B}
			
			% TinyLlama
			\addplot[color=tinyllama, mark=o] coordinates {
				(1,0.9760) (2,0.9377) (3,0.9648) (4,0.9388) (5,0.9784) (6,0.9784) (7,0.9783) (8,0.9651) (9,0.9650) (10,0.9778)
			};
			\addlegendentry{TinyLlama-1.1B-Chat-v1.0}
			
		\end{axis}
	\end{tikzpicture}
	\caption{F1-Score over 10 federated rounds as the cumulative privacy budget $\varepsilon$ is consumed.}
	\label{fig:f1}
\end{figure}
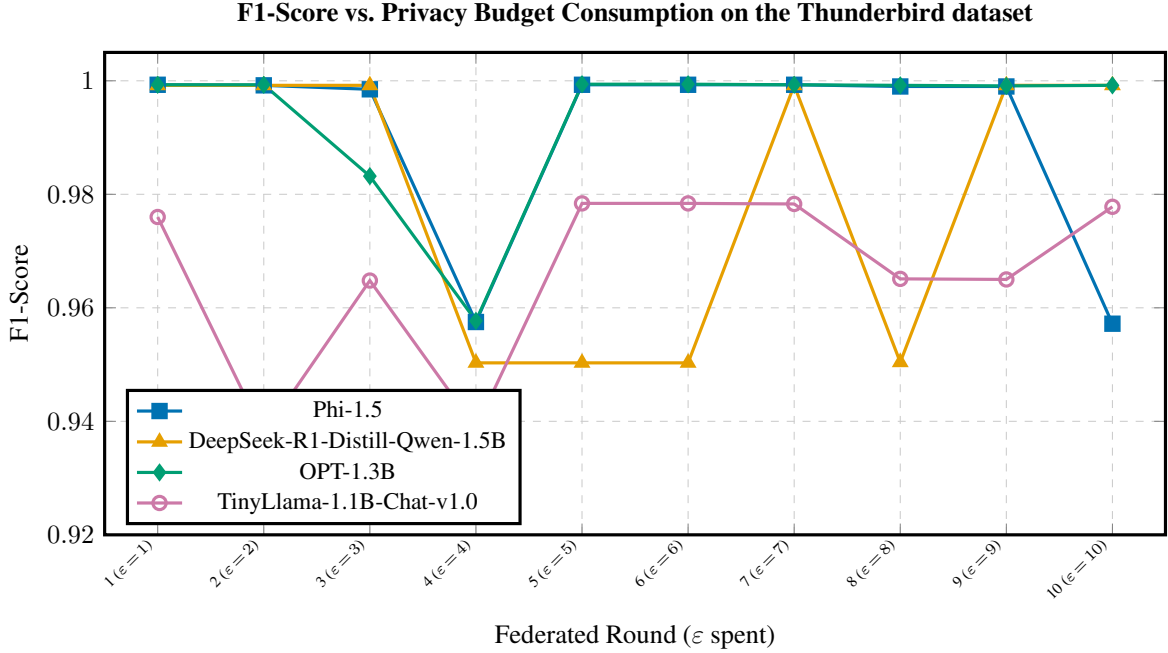

\subsubsection{Privacy Utility Trade-off Analysis}

Figure \ref{fig:f1} illustrates the F1-score versus privacy consumption of the FlogTinyLLM variant for each round. Under the DP framework with targeted  budget, $\varepsilon = 10.0$, the per-round privacy expenditure of $\varepsilon_r = 1.0$ represents a moderate privacy regime. The noise injection magnitude ($\sigma = 0.01$, $\|n\| \approx 3.5-5.1$) is calibrated to the minimal noise setting, prioritizing utility over strict privacy guarantees. Key observations:

\begin{enumerate}
	\item \textbf{DP Noise Impact:} From Table \ref{tab:architecture}, the DP noise norm varies by model due to differing trainable parameter counts. Phi-1.5's larger trainable parameter space (3.15M) results in higher noise magnitude compared to DeepSeek (1.49M).
	
	\item \textbf{Precision Resilience:} All models maintain exceptional precision ($> 0.9994$) despite DP noise, indicating that false positive rates are robustly controlled under differential privacy constraints.
	
	\item \textbf{Recall Sensitivity:} Recall metrics show greater sensitivity to DP noise, with TinyLlama exhibiting the largest recall variance ($\pm 0.0278$), suggesting smaller models may require larger privacy budgets or reduced noise for comparable recall performance.
\end{enumerate}

\subsubsection{Statistical Stability Analysis }
%==============================================================================
Training with differential privacy adds noise, which raises a natural question as to  whether the model performance is stable or unstable. To answer this, we record the accuracy, F1-score, and ROC-AUC at each of the 10 federated  rounds and compute the mean and standard deviation for each model variant. Table~\ref{tab:stability}  shows that
OPT-1.3B has the lowest stability Index at 0.0142. Its accuracy holds at $0.9876 \pm 0.0268$, its F1-score  at $0.9935 \pm 0.0138$, and its ROC-AUC at  $0.9936 \pm 0.0021$. All three metrics stay close to  their means throughout the 10 rounds, indicating that  the DP noise does not push this model off course during training.

Phi-1.5 and TinyLlama fall in the middle, with Stability indices of 0.0165 and 0.0171, respectively. Both models perform well on average but show slightly more variation 
between rounds than OPT-1.3B. TinyLlama has the lowest mean accuracy at $0.9355 \pm 0.0285$, though its round-to-round fluctuation remains moderate.
DeepSeek-R1 shows the widest swings. Its accuracy standard deviation of 0.0441 is the highest among all four models, and its Stability Index of 0.0230 is the 
largest in the table. This suggests that its internal  structure is more affected by the noise that differential privacy introduces at each aggregation step.

\begin{table*}[h]
	\centering
	\caption{Performance stability metrics (Mean $\pm$ Std) across all 10 federated rounds.}
	\label{tab:stability}
	\begin{threeparttable}
		\begin{tabular}{@{}lcccc@{}}
			\toprule
			\textbf{Model} & \textbf{Accuracy} & \textbf{F1-Score} & \textbf{ROC-AUC} & \textbf{Stability}$^\dagger$ \\
			\midrule
			% \rowcolor{lightblue}
			Phi-1.5 & $0.9745 \pm 0.0313$ & $0.9868 \pm 0.0164$ & $0.9941 \pm 0.0017$ & 0.0165 \\
			% \rowcolor{lightorange}
			DeepSeek-R1 & $0.9618 \pm 0.0441$ & $0.9797 \pm 0.0234$ & $0.9971 \pm 0.0014$ & 0.0230 \\
			% \rowcolor{lightgreen}
			OPT-1.3B & $0.9876 \pm 0.0268$ & $0.9935 \pm 0.0138$ & $0.9936 \pm 0.0021$ & \textbf{0.0142} \\
			% \rowcolor{lightpink}
			TinyLlama & $0.9355 \pm 0.0285$ & $0.9660 \pm 0.0155$ & $0.9851 \pm 0.0073$ & 0.0171 \\
			\bottomrule
		\end{tabular}
		\begin{tablenotes}
			\small
			\item[$\dagger$] Stability Index = average of standard deviations across metrics.
		\end{tablenotes}
	\end{threeparttable}
\end{table*}

\begin{figure*}[h]
	\centering
	\begin{tikzpicture}
		\begin{axis}[
			ybar,
			width=0.8\textwidth,
			height=7cm,
			bar width=14pt,
			xlabel={Model},
			ylabel={Standard Deviation},
			symbolic x coords={Phi-1.5, DeepSeek-R1, OPT-1.3B, TinyLlama},
			xtick=data,
			ymin=0, ymax=0.08,
			ytick={0,0.01,0.02,0.03,0.04,0.05,0.06,0.07},
			legend pos=north east,
			legend style={font=\small, fill=white, fill opacity=0.9, draw=gray!50},
			enlarge x limits=0.15,
			nodes near coords,
			nodes near coords style={font=\tiny, rotate=90, anchor=west},
			every node near coord/.append style={yshift=1pt},
			grid=major,
			grid style={line width=0.1pt, draw=gridgray},
			]
			
			\addplot[fill=phi!70, draw=phi!80!black] coordinates {
				(Phi-1.5, 0.0313) (DeepSeek-R1, 0.0441) (OPT-1.3B, 0.0268) (TinyLlama, 0.0285)
			};
			
			\addplot[fill=deepseek!70, draw=deepseek!80!black] coordinates {
				(Phi-1.5, 0.0164) (DeepSeek-R1, 0.0234) (OPT-1.3B, 0.0138) (TinyLlama, 0.0155)
			};
			
			\addplot[fill=opt!70, draw=opt!80!black] coordinates {
				(Phi-1.5, 0.0017) (DeepSeek-R1, 0.0014) (OPT-1.3B, 0.0021) (TinyLlama, 0.0073)
			};
			
			\legend{Accuracy $\sigma$, F1 $\sigma$, AUC $\sigma$}
		\end{axis}
	\end{tikzpicture}
	\caption{Variance decomposition across performance metrics. OPT-1.3B demonstrates the lowest overall variance, indicating the most stable training dynamics under differential privacy.}
	\label{fig:variance}
\end{figure*}
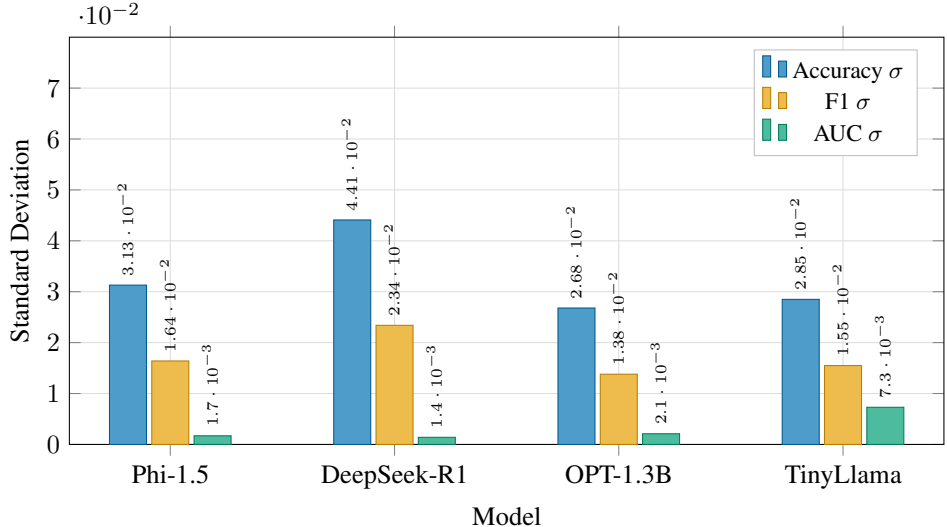

\subsubsection{Computational Efficiency Analysis }
%==============================================================================
Table~\ref{tab:efficiency} shows how long each FlogTinyLLM variant takes to complete the full 10-round federated training on the Thunderbird dataset, and how productively it spends that time. Phi-1.5 finishes fastest at 105.1 hours total, averaging 63.1 minutes per round. It also achieves the highest  efficiency score of $2.53e-5$, which is computed as the final F1-score divided by the total training time in seconds. This means Phi-1.5 delivers more detection quality per unit of compute than any other variant. Its time per parameter is 12.01 milliseconds, the lowest in the table, and reflects the smaller footprint of its trainable LoRA matrices relative to the other variants.

DeepSeek-R1 and OPT-1.3B occupy the middle ground. DeepSeek-R1 takes 156.9 hours with 94.2 minutes per round, while OPT-1.3B takes slightly longer at 165.6 hours and 99.4 minutes per round. Despite being slower, OPT-1.3B spends less time per parameter than DeepSeek-R1, which suggests that OPT-1.3B processes its parameters more efficiently but has more of them to get through. TinyLlama is the most expensive to train. It requires 194.1 hours in total, averaging 116.4 minutes per round, and its time per parameter of 45.49 milliseconds is the highest among all four variants. Its Efficiency score of 1.40e-5 is the lowest, meaning it takes the longest to reach each unit of F1 performance.

\begin{table*}[h]
	\centering
	\caption{Computational efficiency metrics comparison.}
	\label{tab:efficiency}
	\begin{threeparttable}
		\begin{tabular}{@{}lcccc@{}}
			\toprule
			\textbf{Model} & \textbf{Total (h)} & \textbf{Avg/Round (min)} & \textbf{Time/Param (ms)} & \textbf{Efficiency}$^\ddagger$ \\
			\midrule
			% \rowcolor{lightblue}
			Phi-1.5 & 105.1 & 63.1 & 12.01 & \textbf{2.53e-5} \\
			% \rowcolor{lightorange}
			DeepSeek-R1 & 156.9 & 94.2 & 37.81 & 1.77e-5 \\
			% \rowcolor{lightgreen}
			OPT-1.3B & 165.6 & 99.4 & 25.23 & 1.67e-5 \\
			% \rowcolor{lightpink}
			TinyLlama & 194.1 & 116.4 & 45.49 & 1.40e-5 \\
			\bottomrule
		\end{tabular}
		\begin{tablenotes}
			\small
			\item[$\ddagger$] Efficiency = Final F1 / Total Training Time in seconds (higher is better).
		\end{tablenotes}
	\end{threeparttable}
\end{table*}

\begin{figure*}[h]
	\centering
	\begin{tikzpicture}
		\begin{axis}[
			width=0.8\textwidth,
			height=7cm,
			xlabel={Federated Round},
			ylabel={Training Time (×$10^3$ seconds)},
			xmin=0.5, xmax=10.5,
			ymin=10, ymax=140,
			xtick={1,2,3,4,5,6,7,8,9,10},
			legend pos=north east,
			legend style={font=\small, fill=white, fill opacity=0.9, draw=gray!50},
			grid=both,
			grid style={line width=0.1pt, draw=gridgray},
			major grid style={line width=0.2pt, draw=gray!60},
			mark size=2.5pt,
			line width=0.6pt
			]
			
			% Phi-1.5 Time (in thousands) - VERIFIED DATA
			\addplot[color=phi, mark=square*, mark options={fill=phi, scale=1.2}] coordinates {
				(1,55.033) (2,43.802) (3,17.673) (4,15.839) (5,41.081) 
				(6,56.798) (7,72.590) (8,34.813) (9,21.403) (10,19.418)
			};
			\addlegendentry{Phi-1.5}
			
			% DeepSeek Time - VERIFIED DATA
			\addplot[color=deepseek, mark=triangle*, mark options={fill=deepseek, scale=1.3}] coordinates {
				(1,81.420) (2,58.883) (3,29.983) (4,26.932) (5,53.613) 
				(6,79.614) (7,106.771) (8,58.742) (9,36.257) (10,32.695)
			};
			\addlegendentry{DeepSeek-R1}
			
			% OPT-1.3B Time - VERIFIED DATA
			\addplot[color=opt, mark=diamond*, mark options={fill=opt, scale=1.3}] coordinates {
				(1,87.628) (2,74.313) (3,29.927) (4,26.879) (5,69.571) 
				(6,96.072) (7,122.917) (8,58.245) (9,16.080) (10,14.596)
			};
			\addlegendentry{OPT-1.3B}
			
			% TinyLlama Time - VERIFIED DATA
			\addplot[color=tinyllama, mark=o, mark options={fill=tinyllama, scale=1.2}] coordinates {
				(1,112.727) (2,104.013) (3,28.590) (4,25.001) (5,76.591) 
				(6,104.248) (7,129.826) (8,53.370) (9,33.840) (10,30.474)
			};
			\addlegendentry{TinyLlama}
			
			% Annotation for Round 7 peak
			\node[font=\scriptsize, anchor=south west] at (axis cs:7.2,130) {Peak: Round 7};
			
		\end{axis}
	\end{tikzpicture}
	\caption{Per-round training time comparison. Significant variance is observed due to client data size heterogeneity, with Round 7 consistently showing peak computation across all models.}
	\label{fig:training_time}
\end{figure*}
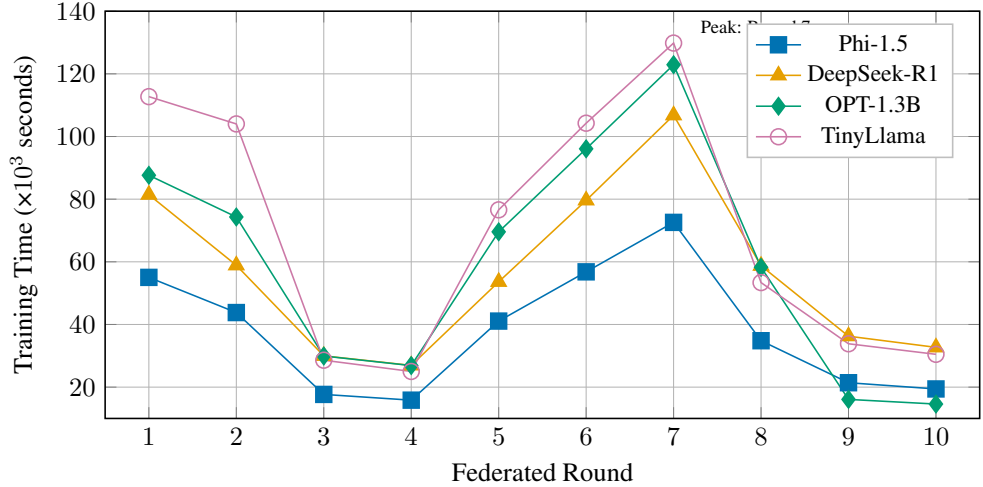

\subsubsection{Comprehensive Model Ranking on Thunderbird}
%==============================================================================
% Define professional color palette
\definecolor{phi15}{RGB}{31, 119, 180}
\definecolor{deepseek}{RGB}{255, 127, 14}
\definecolor{opt13b}{RGB}{44, 160, 44}
\definecolor{tinyllama}{RGB}{214, 39, 40}

% Average Performance Ranking Bar Chart
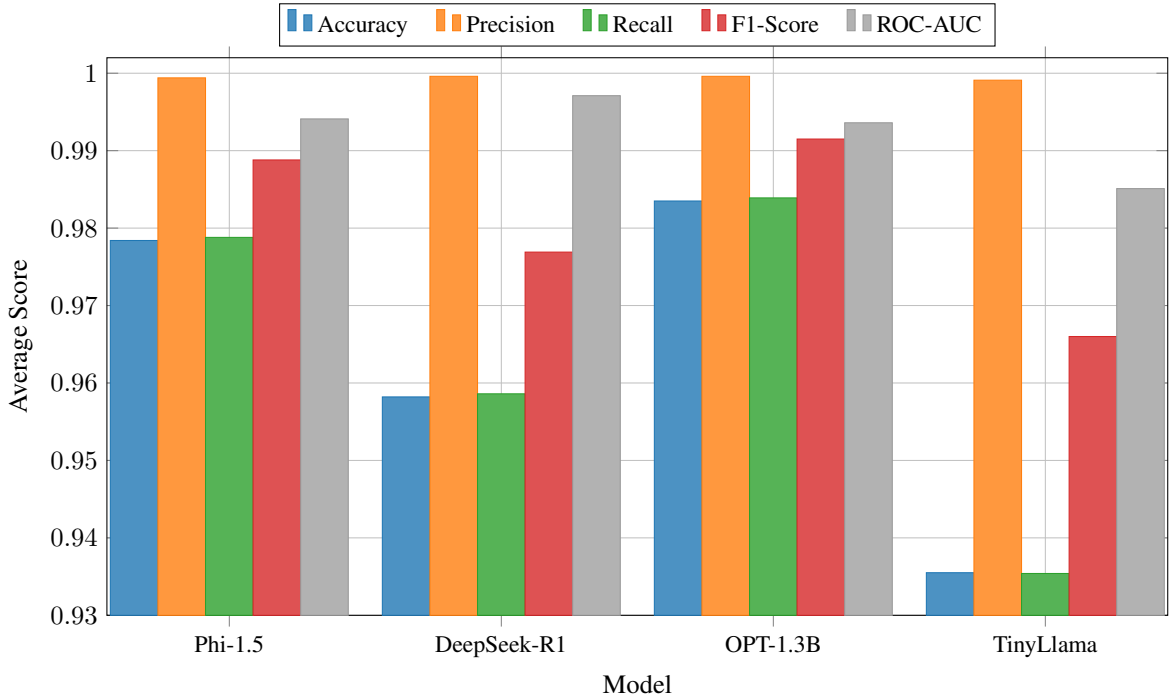
\begin{figure*}[h]
	\centering
	\begin{tikzpicture}
		\begin{axis}[
			width=0.95\textwidth,
			height=9cm,
			ybar=0pt,
			bar width=18pt,
			xlabel={Model},
			ylabel={Average Score},
			ymin=0.93, ymax=1.002,
			symbolic x coords={Phi-1.5, DeepSeek-R1, OPT-1.3B, TinyLlama},
			xtick=data,
			x tick label style={font=\small},
			ytick={0.93, 0.94, 0.95, 0.96, 0.97, 0.98, 0.99, 1.00},
			legend style={at={(0.5,1.02)}, anchor=south, legend columns=5, font=\footnotesize, /tikz/every even column/.append style={column sep=0.3cm}},
			grid=both,
			grid style={line width=0.1pt, draw=gray!30},
			major grid style={line width=0.2pt, draw=gray!50},
			title style={font=\bfseries},
			title={Average Performance Metrics Across All Rounds},
			enlarge x limits=0.15,
			nodes near coords style={font=\tiny, rotate=90, anchor=west},
			]
			
			% Average Accuracy: Phi-1.5=0.9784, DeepSeek-R1=0.9582, OPT-1.3B=0.9835, TinyLlama=0.9355
			\addplot[fill=phi15!80, draw=phi15] coordinates {(Phi-1.5, 0.9784) (DeepSeek-R1, 0.9582) (OPT-1.3B, 0.9835) (TinyLlama, 0.9355)};
			
			% Average Precision: Phi-1.5=0.9994, DeepSeek-R1=0.9996, OPT-1.3B=0.9996, TinyLlama=0.9991
			\addplot[fill=deepseek!80, draw=deepseek] coordinates {(Phi-1.5, 0.9994) (DeepSeek-R1, 0.9996) (OPT-1.3B, 0.9996) (TinyLlama, 0.9991)};
			
			% Average Recall: Phi-1.5=0.9788, DeepSeek-R1=0.9586, OPT-1.3B=0.9839, TinyLlama=0.9354)
			\addplot[fill=opt13b!80, draw=opt13b] coordinates {(Phi-1.5, 0.9788) (DeepSeek-R1, 0.9586) (OPT-1.3B, 0.9839) (TinyLlama, 0.9354)};
			
			% Average F1: Phi-1.5=0.9888, DeepSeek-R1=0.9769, OPT-1.3B=0.9915, TinyLlama=0.9660
			\addplot[fill=tinyllama!80, draw=tinyllama] coordinates {(Phi-1.5, 0.9888) (DeepSeek-R1, 0.9769) (OPT-1.3B, 0.9915) (TinyLlama, 0.9660)};
			
			% Average ROC-AUC: Phi-1.5=0.9941, DeepSeek-R1=0.9971, OPT-1.3B=0.9936, TinyLlama=0.9851
			\addplot[fill=gray!60, draw=gray!80] coordinates {(Phi-1.5, 0.9941) (DeepSeek-R1, 0.9971) (OPT-1.3B, 0.9936) (TinyLlama, 0.9851)};
			
			\legend{Accuracy, Precision, Recall, F1-Score, ROC-AUC}
		\end{axis}
	\end{tikzpicture}
	\caption{Average performance metrics comparison of FlogTinyLLM across all 10 federated rounds. OPT-1.3B achieves the highest average accuracy (0.9835), recall (0.9839), and F1-score (0.9915), while DeepSeek-R1 leads in ROC-AUC (0.9971).}
	\label{fig:avg_performance}
\end{figure*}

\begin{table*}[h]
	\centering
	\caption{Time and Overhead Comparison (Thunderbird Dataset). 
		Overhead = FlogTinyLLM / LogTinyLLM.}
	\label{tab:tradeoff_time}
	\begin{tabular}{lccc}
		\toprule
		Model & LogTinyLLM (h) & FlogTinyLLM  & Time Overhead  \\
		&                &   Time (h)           &   ($\times$)   \\
		\midrule
		microsoft/phi-1.5                  & 17.54 & 105.12 & 5.99 \\
		DeepSeek-R1-Distill-Qwen-1.5B & 16.00  & 156.92 & 9.81 \\
		facebook/opt-1.3b                  & 14.50  & 165.62 & 11.42 \\
		TinyLlama-1.1B-Chat-v1.0           & 12.20 & 194.08 & 15.91 \\
		\bottomrule
	\end{tabular}
\end{table*}

Table \ref{tab:tradeoff_time} shows the computational time requirements between LogTinyLLM and FlogTinyLLM across the four TinyLLMs on the Thunderbird dataset.
The table shows variation in time overhead across the different frameworks. Phi-1.5 shows the best time efficiency profile, requiring 17.54 hours for LogTinyLLM and 105.12 hours for FlogTinyLLM, yielding the lowest time overhead of 5.99$\times$. In contrast, TinyLlama-1.1B exhibits the highest overhead at 15.91$\times$, with LogTinyLLM completing in 12.20 hours compared to 194.08 hours for FlogTinyLLM.
An inverse relationship is seen between baseline execution time and overhead ratio. TinyLlama-1.1B achieves the fastest LogTinyLLM execution  of 12.20 hours yet incurs the greatest overhead multiplication. Conversely, Phi-1.5, despite having the longest LogTinyLLM runtime of 17.54 hours, has the lowest overhead factor.
The deepseek 1.5B model represents an intermediary efficiency with 16.00 hours for LogTinyLLM and 156.92 hours for FlogTinyLLM, resulting in a 9.81$\times$ overhead. Similarly, Opt-1.3b requires 14.50 hours and 165.62 hours, respectively, producing an 11.42$\times$ overhead.
These indicate that FlogTinyLLM introduces higher computational costs, ranging from approximately 6$\times$ to nearly 16$\times$ the baseline time in LogTinyLLM, suggesting important trade-offs between the two approaches\cite{ocansey2025logtinyllm}.

\subsubsection{Comparative Analysis of LogTinyLLM and FlogTinyLLM on the Thunderbird Dataset}
\definecolor{federated}{RGB}{31, 119, 180}
\definecolor{centralized}{RGB}{255, 127, 14}
\definecolor{phi15}{RGB}{44, 160, 44}
\definecolor{deepseek}{RGB}{214, 39, 40}
\definecolor{opt13b}{RGB}{148, 103, 189}
\definecolor{tinyllama}{RGB}{140, 86, 75}
\definecolor{positive}{RGB}{39, 174, 96}
\definecolor{negative}{RGB}{231, 76, 60}
\definecolor{headerblue}{RGB}{41, 128, 185}

%=============================================================================
% COMPREHENSIVE COMPARISON TABLE
%=============================================================================
\begin{table*}[h]
	\centering
	\caption{Performance Comparison: Centralized LogTinyLLM vs. Federated FlogTinyLLM }
	\label{tab:comparison}
	\footnotesize
	\begin{tabular}{@{}l l c c c c@{}}
		\toprule
		\textbf{Model} & \textbf{Framework} & \textbf{Accuracy} & \textbf{Precision} & \textbf{Recall} & \textbf{F1-Score} \\
		\midrule
		\multirow{3}{*}{\textbf{Phi-1.5}} 
		& LogTinyLLM & 0.9868 & 0.9847 & 0.9586 & 0.9842 \\
		& FlogTinyLLM (Avg) & 0.9825 & 0.9994 & 0.9828 & 0.9908 \\
		& \cellcolor{gray!10}$\Delta$ (Fed - Cent) & \cellcolor{gray!10}\textcolor{negative}{-0.0043} & \cellcolor{gray!10}\textcolor{positive}{+0.0147} & \cellcolor{gray!10}\textcolor{positive}{+0.0242} & \cellcolor{gray!10}\textcolor{positive}{+0.0066} \\
		\midrule
		\multirow{3}{*}{\textbf{DeepSeek-R1}} 
		& LogTinyLLM & 0.9883 & 0.9877 & 0.9605 & 0.9856 \\
		& FlogTinyLLM (Avg) & 0.9618 & 0.9996 & 0.9615 & 0.9797 \\
		& \cellcolor{gray!10}$\Delta$ (Fed - Cent) & \cellcolor{gray!10}\textcolor{negative}{-0.0265} & \cellcolor{gray!10}\textcolor{positive}{+0.0119} & \cellcolor{gray!10}\textcolor{positive}{+0.0010} & \cellcolor{gray!10}\textcolor{negative}{-0.0060} \\
		\midrule
		\multirow{3}{*}{\textbf{OPT-1.3B}} 
		& LogTinyLLM & 0.9873 & 0.9857 & 0.9693 & 0.9848 \\
		& FlogTinyLLM (Avg) & 0.9876 & 0.9996 & 0.9879 & 0.9935 \\
		& \cellcolor{gray!10}$\Delta$ (Fed - Cent) & \cellcolor{gray!10}\textcolor{positive}{+0.0003} & \cellcolor{gray!10}\textcolor{positive}{+0.0139} & \cellcolor{gray!10}\textcolor{positive}{+0.0186} & \cellcolor{gray!10}\textcolor{positive}{+0.0087} \\
		\midrule
		\multirow{3}{*}{\textbf{TinyLlama}} 
		& LogTinyLLM & 0.9865 & 0.9842 & 0.9491 & 0.9835 \\
		& FlogTinyLLM (Avg) & 0.9355 & 0.9991 & 0.9354 & 0.9660 \\
		& \cellcolor{gray!10}$\Delta$ (Fed - Cent) & \cellcolor{gray!10}\textcolor{negative}{-0.0510} & \cellcolor{gray!10}\textcolor{positive}{+0.0149} & \cellcolor{gray!10}\textcolor{negative}{-0.0137} & \cellcolor{gray!10}\textcolor{negative}{-0.0175} \\
		\bottomrule
	\end{tabular}
	
	\vspace{0.3cm}
	\footnotesize
	\textit{Note:} Federated values represent the average across 10 federated rounds. \\
	\textcolor{positive}{Green (+)} indicates federated outperforms centralized; \textcolor{negative}{Red (-)} indicates centralized outperforms federated.
\end{table*}

Table \ref{tab:comparison} shows the performance outlook of the centralized LogTinyLLM and the federated FlogTinyLLM on the Thunderbird dataset. The Phi-1.5 model shows good performance on FlogTinyLLM, with increases in Precision by 0.0147, Recall by 0.0242, and F1-Score by 0.0066, while marginally decreasing in  accuracy by 0.0043. Phi-1.5 0.9828 Recall under FlogTinyLLM, compared to 0.9586 under LogTinyLLM, indicates better log sequence classification in the federated framework. The DeepSeek-R1 also exhibited a trade-off between accuracy and specificity, showing a 0.0265 reduction in accuracy and a 0.0060 reduction in F1-score under FlogTinyLLM. However, precision reached a near-perfect value of 0.9996. From Table \ref{tab:comparison}, OPT-1.3B is identified as the most stable candidate under FlogTinyLLM, achieving superior performance across all metrics compared to the centralized LogTinyLLM. This suggests a unique structural compatibility between OPT-1.3B's weight distribution and the FlogTinyLLM architecture. TinyLlama showed  a sharp sensitivity to the federated framework. While precision improved by 0.0149, the model suffered regressions in Accuracy by 0.0510, Recall by 0.0137, and F1-Score by 0.0175. The table shows that FlogTinyLLM consistently improves Precision across all architectures. However, architectural stability varies. OPT-1.3B and Phi-1.5 are the most viable candidates for the FlogTinyLLM framework compared to the LogTinyLLM architecture.

%=============================================================================
% GROUPED BAR CHART - SIDE BY SIDE COMPARISON
%=============================================================================
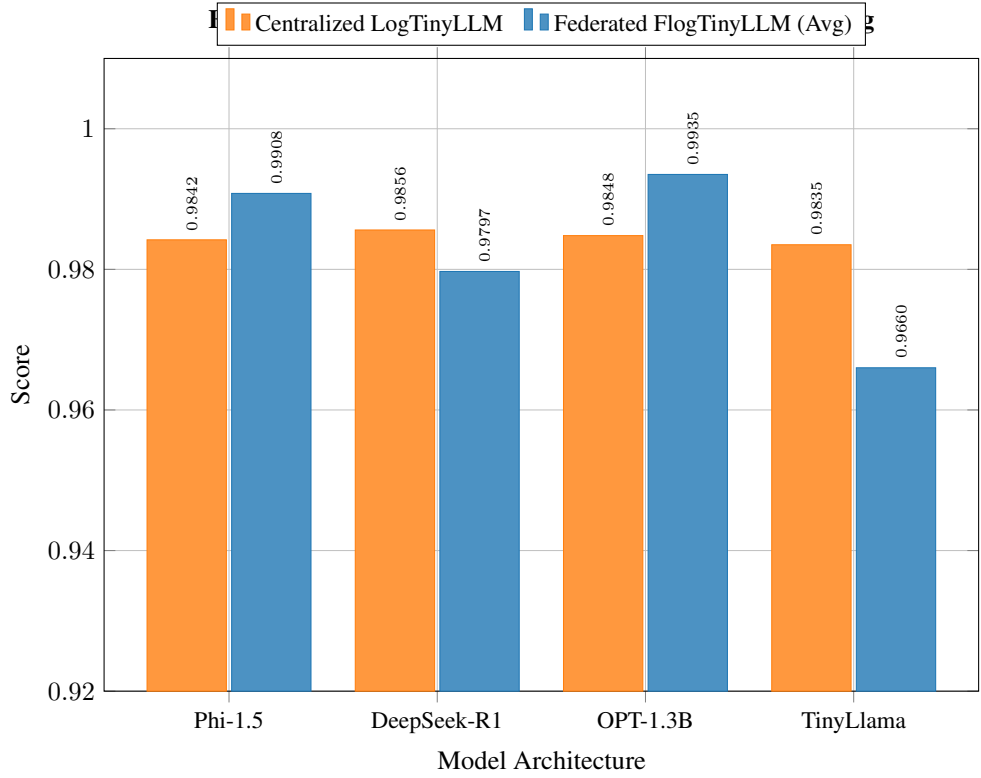
\begin{figure*}[h]
	\centering
	\begin{tikzpicture}
		\begin{axis}[
			width=0.8\textwidth,
			height=10cm,
			ybar=2pt,
			bar width=30pt,
			xlabel={Model Architecture},
			ylabel={Score},
			ymin=0.92, ymax=1.01,
			symbolic x coords={Phi-1.5, DeepSeek-R1, OPT-1.3B, TinyLlama},
			xtick=data,
			x tick label style={font=\small},
			ytick={0.92, 0.94, 0.96, 0.98, 1.00},
			legend style={at={(0.5,1.02)}, anchor=south, legend columns=4, font=\footnotesize, /tikz/every even column/.append style={column sep=0.2cm}},
			grid=both,
			grid style={line width=0.1pt, draw=gray!30},
			major grid style={line width=0.2pt, draw=gray!50},
			title style={font=\bfseries},
			title={F1-Score Comparison: Centralized vs. Federated Learning},
			enlarge x limits=0.2,
			nodes near coords,
			nodes near coords style={font=\tiny, rotate=90, anchor=west, xshift=0pt, yshift=-2pt},
			every node near coord/.append style={/pgf/number format/.cd, fixed, fixed zerofill, precision=4},
			]
			
			% Centralized F1-Scores (exact from uploaded table)
			\addplot[fill=centralized!80, draw=centralized] coordinates {
				(Phi-1.5, 0.9842) 
				(DeepSeek-R1, 0.9856) 
				(OPT-1.3B, 0.9848) 
				(TinyLlama, 0.9835)
			};
			
			% Federated F1-Scores (exact calculated averages)
			\addplot[fill=federated!80, draw=federated] coordinates {
				(Phi-1.5, 0.9908) 
				(DeepSeek-R1, 0.9797) 
				(OPT-1.3B, 0.9935) 
				(TinyLlama, 0.9660)
			};
			
			\legend{Centralized LogTinyLLM, Federated FlogTinyLLM (Avg)}
		\end{axis}
	\end{tikzpicture}
	\caption{F1-Score comparison between LogTinyLLM and FlogTinyLLM frameworks. OPT-1.3B shows the best improvement under FlogTinyLLM by 0.0087, achieving the highest overall F1-Score. Phi-1.5 also sees an F1-Score benefit from the FlogTinyLLM framework.}
	\label{fig:f1_comparison}
\end{figure*}

\begin{table*}[h]
	\centering
	\caption{LogTinyLLM vs. FlogTinyLLM Framework Analysis on the Thunderbird Dataset }
	\label{tab:key_findings}
	\footnotesize
	\begin{tabular}{lll}
		\toprule
		\textbf{Metric} & \textbf{Centralized Strength} & \textbf{Federated Strength} \\
		\midrule
		\textbf{Accuracy} & 3 of 4 models higher & OPT-1.3B only (+0.0003) \\
		& Best: DeepSeek-R1 (0.9883) & Best: OPT-1.3B (0.9876) \\
		& Range: 0.9865--0.9883 & Range: 0.9355--0.9876 \\
		\midrule
		\textbf{Precision} & -- & \textbf{All 4 models higher} \\
		& Best: DeepSeek-R1 (0.9877) & Best: OPT-1.3B/DeepSeek (0.9996) \\
		& Range: 0.9842--0.9877 & Range: 0.9991--0.9996 \\
		\midrule
		\textbf{Recall} & TinyLlama only & 3 of 4 models higher \\
		& Best: OPT-1.3B (0.9693) & Best: OPT-1.3B (0.9879) \\
		& Gain: Phi +0.0242, OPT +0.0186 & DeepSeek +0.0010 \\
		\midrule
		\textbf{F1-Score} & DeepSeek-R1, TinyLlama & Phi-1.5, OPT-1.3B \\
		& Best: DeepSeek-R1 (0.9856) & \textbf{Best: OPT-1.3B (0.9935)} \\
		& & Highest overall F1 \\
		\midrule
		\textbf{Best Overall} & -- & \textbf{OPT-1.3B Federated} \\
		& & Only model with all positive $\Delta$ \\
		\bottomrule
	\end{tabular}
\end{table*}

Table \ref{tab:key_findings} presents a comparative analysis of LogTinyLLM and FlogTinyLLM approaches across metrics on the Thunderbird dataset.
LogTinyLLM demonstrates superior accuracy performance, with 3 of 4 models achieving higher scores. DeepSeek-R1 achieves the best centralized accuracy of 0.9883, with scores ranging from 0.9865 to 0.9883.  FlogTinyLLM shows a narrower advantage, with only OPT-1.3B achieving a marginal improvement of 0.0003. FlogTinyLLM accuracy ranges more broadly from 0.9355 to 0.9876.
FlogTinyLLM shows a clear advantage in Precision, with all four models achieving higher precision, ranging from 0.9991 to 0.9996; OPT-1.3B and DeepSeek both achieving the best scores. However, this gain in precision comes with a trade-off in accuracy. The results illustrate that while precision climbs, notably enhancing FlogTinyLLM's performance, there is a modest dip in accuracy, with LogTinyLLM maintaining slightly higher accuracy in certain configurations. LogTinyLLM precision ranges from 0.9842 to 0.9877, with DeepSeek-R1 achieving the highest at 0.9877.
Results are mixed in Recall. Among the models, FlogTinyLLM shows higher recall in 3 out of 4 configurations, whereas only TinyLlama favors the LogTinyLLM. When comparing the frameworks, FlogTinyLLM achieves the best recall score with OPT-1.3B at 0.9879. In contrast, the best score for LogTinyLLM is 0.9693.
There are almost even performances between FlogTinyLLM and LogTinyLLM on F1-Score, with DeepSeek-R1 and TinyLlama favoring LogTinyLLM training, while Phi-1.5 and OPT-1.3B favoring FlogTinyLLM training. Notably, OPT-1.3B achieves the highest overall F1-score of 0.9935 under the FlogTinyLLM framework, surpassing the best LogTinyLLM F1-score of 0.9856 in DeepSeek-R1. This higher F1-score of OPT-1.3B indicates its suitability for applications where balanced precision and recall are critical. 
Overall, OPT-1.3B under the FlogTinyLLM framework is the best performer, being the only model configuration with all positive performance differences across metrics.

\begin{figure*}[h]
\centering
\begin{tikzpicture}
\begin{axis}[
    width=0.8\textwidth,
    height=7cm,
    ybar,
    bar width=10pt,
    xlabel={Model Architecture},
    ylabel={Performance Delta},
    ymin=-0.06, ymax=0.03,
    symbolic x coords={Phi15, DeepSeekR1, OPT13B, TinyLlama},
    xtick=data,
    xticklabels={Phi-1.5, DeepSeek-R1, OPT-1.3B, TinyLlama},
    x tick label style={font=\small, rotate=15, anchor=east},
    ytick={-0.05,-0.03,-0.01,0,0.01,0.02},
    yticklabel style={/pgf/number format/fixed},
    legend style={
        at={(0.5,1.05)},
        anchor=south,
        legend columns=2,
        font=\footnotesize
    },
    grid=both,
    grid style={gray!20},
    enlarge x limits=0.2
]

% Accuracy
\addplot+[ybar, fill=blue!60] coordinates {
    (Phi15,-0.0043)
    (DeepSeekR1,-0.0265)
    (OPT13B,0.0003)
    (TinyLlama,-0.0510)
};

% Precision
\addplot+[ybar, fill=orange!70] coordinates {
    (Phi15,0.0147)
    (DeepSeekR1,0.0119)
    (OPT13B,0.0139)
    (TinyLlama,0.0149)
};

% Recall
\addplot+[ybar, fill=green!60] coordinates {
    (Phi15,0.0242)
    (DeepSeekR1,0.0010)
    (OPT13B,0.0186)
    (TinyLlama,-0.0137)
};

% F1
\addplot+[ybar, fill=red!60] coordinates {
    (Phi15,0.0066)
    (DeepSeekR1,-0.0060)
    (OPT13B,0.0087)
    (TinyLlama,-0.0175)
};

\legend{$\Delta$ Accuracy, $\Delta$ Precision, $\Delta$ Recall, $\Delta$ F1}

\end{axis}
\end{tikzpicture}
\caption{Performance delta analysis on the Thunderbird dataset.}
\label{fig:delta_analysis}
\end{figure*}
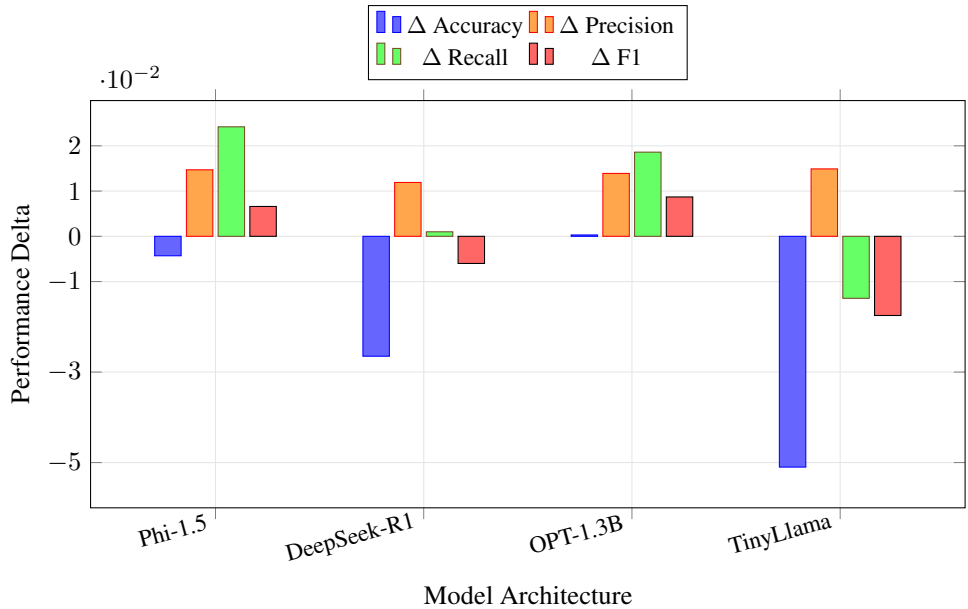

Figure~\ref{fig:delta_analysis} presents a bar chart of performance deltas ($\Delta = \text{FlogTinyLLM} - \text{LogTinyLLM}$) across the four model architectures on the Thunderbird dataset, covering accuracy, precision, recall, and F1-score. The four models have positive deltas in precision under the federated setting. Recall improves for three of the four models: Phi-1.5 ($+0.0242$), OPT-1.3B ($+0.0186$), and DeepSeek-R1 ($+0.0010$ while TinyLlama incurs a slight regression ($-0.0137$). F1-score gains are observed for Phi-1.5 ($+0.0066$) and OPT-1.3B ($+0.0087$), whereas DeepSeek-R1 and TinyLlama experience modest declines of $-0.0060$ and $-0.0175$, respectively. Accuracy is the most adversely affected metric: three models exhibit negative deltas, with TinyLlama recording the largest drop ($-0.0510$) and OPT-1.3B remaining nearly neutral ($+0.0003$).

\begin{figure*}[h]
	\centering
	\begin{tikzpicture}
		\begin{axis}[
			width=0.85\linewidth,
			height=8cm,
			xlabel={Communication Rounds},
			ylabel={Accuracy},
			grid=major,
			grid style={dashed, gray!30},
			legend pos=south east,
			legend style={
				nodes={scale=0.75, transform shape},
				fill=white,
				fill opacity=0.9,
				draw opacity=1,
				text opacity=1,
				draw=black!50,
				rounded corners=2pt
			},
			mark size=1.5pt,
			xmin=0, xmax=21,
			xtick={1,5,10,15,20},
			ymin=0.60, ymax=1.00,
			ytick={0.60,0.70,0.80,0.90,1.00},
			tick label style={font=\small},
			label style={font=\normalsize},
			line width=0.5pt,
			axis line style={-stealth},
			title={FlogTinyLLM Accuracy on BGL Dataset Over 20 Rounds}
			]
			\addplot[
			color=blue,
			mark=square*,
			thick,
			line width=1.5pt
			] coordinates {
				(1,0.9507)
				(2,0.9521)
				(3,0.9666)
				(4,0.9521)
				(5,0.9521)
				(6,0.9521)
				(7,0.9681)
				(8,0.9681)
				(9,0.9681)
				(10,0.9231)
				(11,0.9521)
				(12,0.9521)
				(13,0.9681)
				(14,0.9681)
				(15,0.9681)
				(16,0.9521)
				(17,0.9521)
				(18,0.9521)
				(19,0.9521)
				(20,0.9681)
			};
			\addlegendentry{DeepSeek-R1-Distill-Qwen-1.5B}
			
			\addplot[
			color=red,
			mark=triangle*,
			thick,
			line width=1.5pt
			] coordinates {
				(1,0.9216)
				(2,0.9289)
				(3,0.9332)
				(4,0.9332)
				(5,0.9332)
				(6,0.9332)
				(7,0.9332)
				(8,0.9332)
				(9,0.9332)
				(10,0.9332)
				(11,0.9332)
				(12,0.9332)
				(13,0.9332)
				(14,0.9332)
				(15,0.9332)
				(16,0.9332)
				(17,0.9332)
				(18,0.9332)
				(19,0.9332)
				(20,0.9332)
			};
			\addlegendentry{facebook/opt-1.3b}
			
			\addplot[
			color=green!60!black,
			mark=pentagon*,
			thick,
			line width=1.5pt
			] coordinates {
				(1,0.9419)
				(2,0.9492)
				(3,0.9332)
				(4,0.9332)
				(5,0.9492)
				(6,0.9492)
				(7,0.9492)
				(8,0.9332)
				(9,0.9332)
				(10,0.9492)
				(11,0.9492)
				(12,0.9332)
				(13,0.9492)
				(14,0.9492)
				(15,0.9492)
				(16,0.9492)
				(17,0.9492)
				(18,0.9492)
				(19,0.9492)
				(20,0.9492)
			};
			\addlegendentry{microsoft/phi-1.5}
			
			\addplot[
			color=orange,
			mark=diamond*,
			thick,
			line width=1.5pt
			] coordinates {
				(1,0.9463)
				(2,0.9303)
				(3,0.9303)
				(4,0.9332)
				(5,0.9332)
				(6,0.9332)
				(7,0.9492)
				(8,0.9332)
				(9,0.9332)
				(10,0.9332)
				(11,0.9332)
				(12,0.9332)
				(13,0.9332)
				(14,0.9332)
				(15,0.9492)
				(16,0.9332)
				(17,0.9492)
				(18,0.9492)
				(19,0.9332)
				(20,0.9332)
			};
			\addlegendentry{TinyLlama-1.1B}
		\end{axis}
	\end{tikzpicture}
	\caption{FlogTinyLLM  accuracy over 20 communication rounds on the BGL dataset. DeepSeek-R1 demonstrates superior performance with peak accuracy of 0.9681, while Opt-1.3B exhibits the most stable convergence pattern at 0.9332 from round 3 onwards.}
	\label{fig:fed_learning_acc}
\end{figure*}
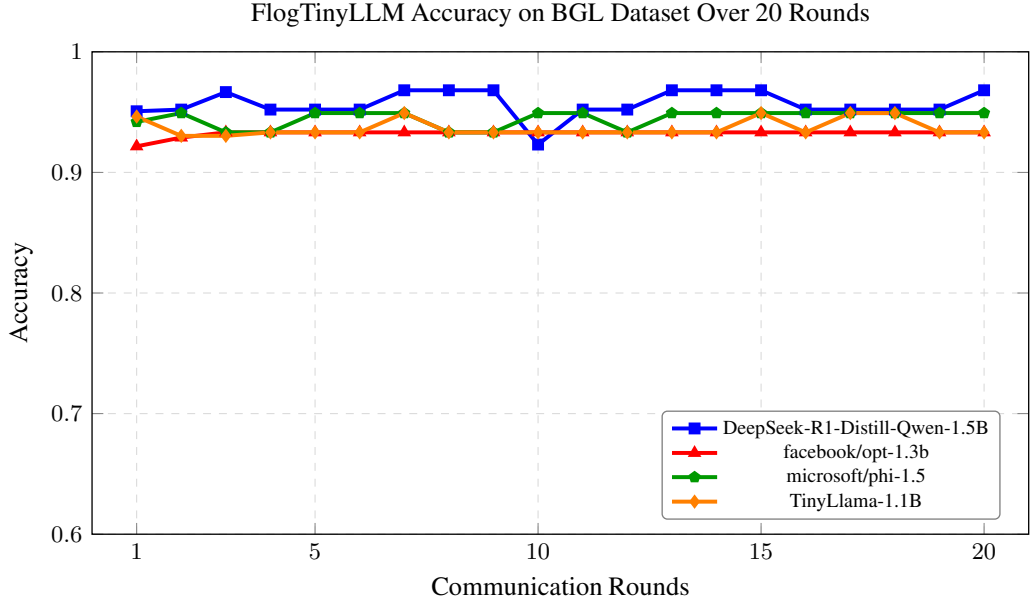

Figure \ref{fig:fed_learning_acc}  and Table \ref{tab:best_performance1} show the performance of FlogTinyLLM accuracy on the BGL dataset over 20 communication rounds. \textbf{DeepSeek-R1} demonstrates the highest overall performance, achieving a peak accuracy of 0.9681 across several rounds (7 to 9, 13 to 15, and 20). Although some variability is observed, including a decrease to 0.9231 at round 10, the model consistently returns to high performance levels. The mean accuracy across all rounds is 0.9580, with a standard deviation of 0.0118. \textbf{Opt-1.3B} demonstrates the most stable performance, converging to an accuracy of 0.9332 by round 3 and maintaining this level consistently through round 20. The low standard deviation of 0.0034 indicates minimal variance and highly predictable behavior throughout the federated learning process. \textbf{Phi-1.5} demonstrates moderate performance, achieving a peak accuracy of 0.9492 and a mean accuracy of 0.9443. The model fluctuates between 0.9332 and 0.9492 across rounds, with a standard deviation of 0.0067, indicating relatively stable but not entirely consistent performance. \textbf{TinyLlama-1.1B} exhibits the most variable performance among the four models, with accuracy ranging from 0.9303 to 0.9492 throughout the training process.

\begin{table*}[h]
	\centering
	\caption{FlogTinyLLM F1-Score on BGL Dataset Over 20 Communication Rounds}
	\label{tab:f1_federated}
	\renewcommand{\arraystretch}{1.15}
	\begin{tabular}{c cccc}
		\toprule
		\textbf{Round} & \textbf{DeepSeek-R1-Distill-Qwen-1.5B} & \textbf{opt-1.3b} & \textbf{phi-1.5} & \textbf{TinyLlama-1.1B} \\
		\midrule
		1  & 0.8426 & 0.7404 & 0.7938 & 0.8083 \\
		2  & 0.8465 & 0.7610 & 0.8168 & 0.7647 \\
		3  & 0.8878 & 0.7723 & 0.7723 & 0.7647 \\
		4  & 0.8465 & 0.7723 & 0.7723 & 0.7723 \\
		5  & 0.8465 & 0.8723 & 0.8768 & 0.7723 \\
		6  & 0.8465 & 0.7923 & 0.8168 & 0.7723 \\
		7  & 0.8922 & 0.7723 & 0.8168 & 0.8168 \\
		8  & 0.8922 & 0.8723 & 0.7723 & 0.7723 \\
		9  & 0.8922 & 0.8123 & 0.8723 & 0.7723 \\
		10 & 0.7764 & 0.7723 & 0.8168 & 0.7723 \\
		\midrule
		11 & 0.8465 & 0.7723 & 0.8768 & 0.7723 \\
		12 & 0.8465 & 0.7723 & 0.7723 & 0.7723 \\
		13 & 0.8922 & 0.7723 & 0.8168 & 0.7723 \\
		14 & 0.8922 & 0.9023 & 0.8168 & 0.7723 \\
		15 & 0.8922 & 0.8723 & 0.8568 & 0.8168 \\
		16 & 0.8465 & 0.8023 & 0.8168 & 0.7723 \\
		17 & 0.8465 & 0.8223 & 0.8168 & 0.8168 \\
		18 & 0.8465 & 0.7723 & 0.9168 & 0.8168 \\
		19 & 0.8465 & 0.9023 & 0.8868 & 0.7723 \\
		20 & 0.8922 & 0.7723 & 0.8168 & 0.7723 \\
		\bottomrule
	\end{tabular}
\end{table*}

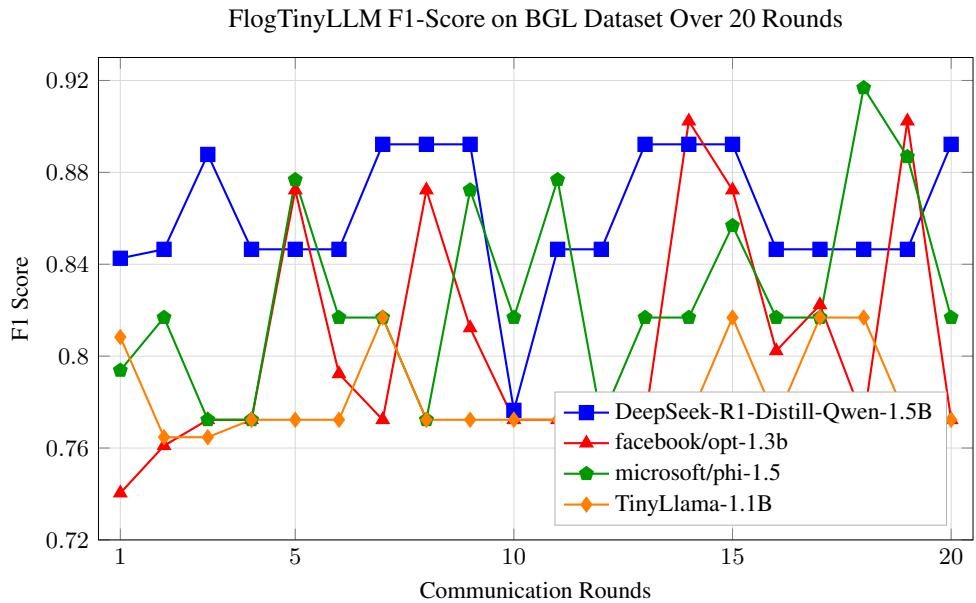
\begin{figure*}[h]
	\centering
	\begin{tikzpicture}
		\begin{axis}[
			width=0.8\linewidth,
			height=8cm,
			xlabel={Communication Rounds},
			ylabel={F1 Score},
			xlabel style={font=\small},
			ylabel style={font=\small},
			grid=major,
			grid style={line width=0.1pt, draw=gray!30},
			legend pos=south east,
			legend style={
				font=\footnotesize,
				cells={anchor=west},
				draw=black!30,
				fill=white,
				fill opacity=0.9,
				text opacity=1
			},
			mark size=2.5pt,
			line width=0.2pt,
			xmin=0.5, xmax=20.5,
			ymin=0.72, ymax=0.93,
			xtick={1,5,10,15,20},
			ytick={0.72,0.76,0.80,0.84,0.88,0.92},
			tick label style={font=\small},
			every axis plot/.append style={thick},
			title={FlogTinyLLM F1-Score on BGL Dataset Over 20 Rounds},
			]
			
			% DeepSeek-R1-Distill-Qwen-1.5B
			\addplot[
			color=blue,
			mark=square*,
			] coordinates {
				(1,0.8426) (2,0.8465) (3,0.8878) (4,0.8465) (5,0.8465)
				(6,0.8465) (7,0.8922) (8,0.8922) (9,0.8922) (10,0.7764)
				(11,0.8465) (12,0.8465) (13,0.8922) (14,0.8922) (15,0.8922)
				(16,0.8465) (17,0.8465) (18,0.8465) (19,0.8465) (20,0.8922)
			};
			\addlegendentry{DeepSeek-R1-Distill-Qwen-1.5B}
			
			% facebook/opt-1.3b
			\addplot[
			color=red,
			mark=triangle*,
			] coordinates {
				(1,0.7404) (2,0.7610) (3,0.7723) (4,0.7723) (5,0.8723)
				(6,0.7923) (7,0.7723) (8,0.8723) (9,0.8123) (10,0.7723)
				(11,0.7723) (12,0.7723) (13,0.7723) (14,0.9023) (15,0.8723)
				(16,0.8023) (17,0.8223) (18,0.7723) (19,0.9023) (20,0.7723)
			};
			\addlegendentry{facebook/opt-1.3b}
			
			% microsoft/phi-1_5
			\addplot[
			color=green!60!black,
			mark=pentagon*,
			] coordinates {
				(1,0.7938) (2,0.8168) (3,0.7723) (4,0.7723) (5,0.8768)
				(6,0.8168) (7,0.8168) (8,0.7723) (9,0.8723) (10,0.8168)
				(11,0.8768) (12,0.7723) (13,0.8168) (14,0.8168) (15,0.8568)
				(16,0.8168) (17,0.8168) (18,0.9168) (19,0.8868) (20,0.8168)
			};
			\addlegendentry{microsoft/phi-1.5}
			
			% TinyLlama-1.1B
			\addplot[
			color=orange,
			mark=diamond*,
			] coordinates {
				(1,0.8083) (2,0.7647) (3,0.7647) (4,0.7723) (5,0.7723)
				(6,0.7723) (7,0.8168) (8,0.7723) (9,0.7723) (10,0.7723)
				(11,0.7723) (12,0.7723) (13,0.7723) (14,0.7723) (15,0.8168)
				(16,0.7723) (17,0.8168) (18,0.8168) (19,0.7723) (20,0.7723)
			};
			\addlegendentry{TinyLlama-1.1B}
			
		\end{axis}
	\end{tikzpicture}
	\caption{FlogTinyLLM performance over 20 communication rounds. F1 scores show varying convergence patterns with DeepSeek-R1 achieving the most stable high performance.}
	\label{fig:f1_federated}
\end{figure*}

Across the 20 communication rounds from Table \ref{tab:f1_federated} and Figure \ref{fig:f1_federated}, DeepSeek-R1-Distill-Qwen-1.5B shows the strongest overall performance with a mean F1 of approximately 0.8609 and the most consistent behavior, clustering around 0.8465 and 0.8922 with only a single dip to 0.7764 at round 10. Phi-1.5 follows with a mean of approximately 0.8260 and produces the highest single-round score of any model (0.9168 at round 18), though it fluctuates between 0.7723 and that peak. Opt-1.3b averages approximately 0.8051 and is the most volatile, ranging from 0.7404 at round 1 to 0.9023 at rounds 14 and 19. TinyLlama-1.1B is the weakest performer, averaging approximately 0.7822, though its narrow range makes it relatively predictable at that lower level. Overall, DeepSeek-R1-Distill-Qwen-1.5B offers the best balance of high F1 and round-to-round stability among the four models evaluated on the BGL dataset.

\definecolor{color1}{RGB}{31,119,180}   % Professional blue
\definecolor{color2}{RGB}{214,39,40}    % Muted red
\definecolor{color3}{RGB}{44,160,44}    % Forest green
\definecolor{color4}{RGB}{255,127,14}   % Warm orange

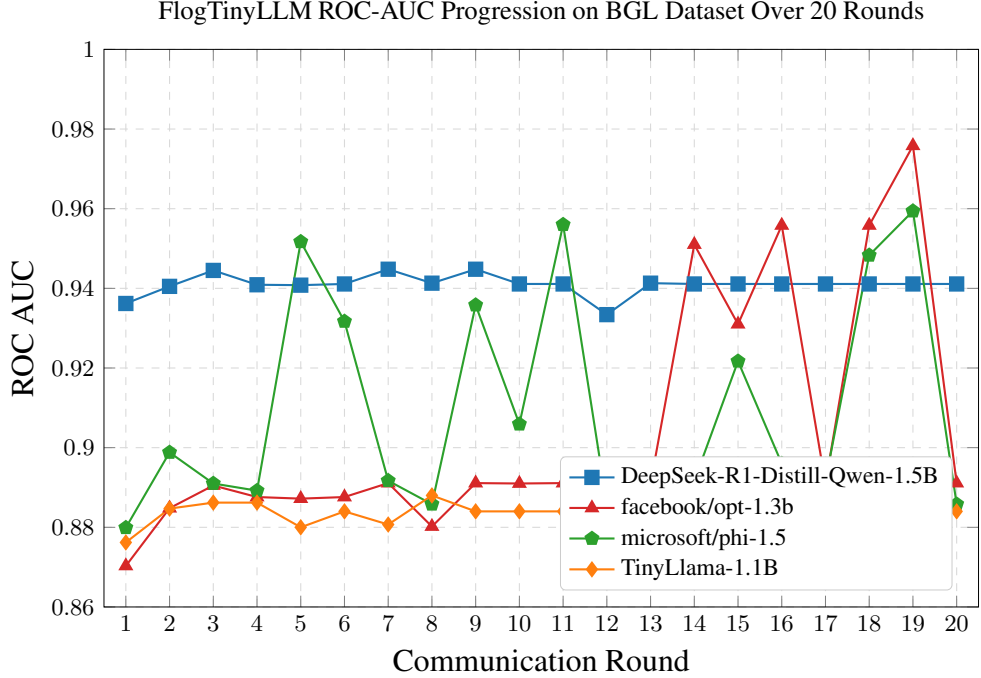
\begin{figure*}[h]
	\centering
	\begin{tikzpicture}
		\begin{axis}[
			width=0.8\textwidth,
			height=9cm,
			xlabel={Communication Round},
			ylabel={ROC AUC},
			xlabel style={font=\large},
			ylabel style={font=\large},
			grid=major,
			grid style={dashed,gray!30},
			legend pos=south east,
			legend style={
				font=\small,
				draw=black!20,
				fill=white,
				fill opacity=0.9,
				text opacity=1,
				rounded corners=2pt,
				cells={anchor=west}
			},
			mark size=2.5pt,
			line width=0.2pt,
			xmin=0.5, xmax=20.5,
			xtick={1,2,3,4,5,6,7,8,9,10,11,12,13,14,15,16,17,18,19,20},
			ymin=0.86, ymax=1.00,
			ytick={0.86,0.88,0.90,0.92,0.94,0.96,0.98,1.00},
			tick label style={font=\footnotesize},
			every axis plot/.append style={thick},
			title={FlogTinyLLM ROC-AUC Progression on BGL Dataset Over 20 Rounds},
			]
			% DeepSeek model
			\addplot[
			color=color1,
			mark=square*,
			mark options={solid,fill=color1}
			] coordinates {
				(1,0.9362) (2,0.9405) (3,0.9445) (4,0.9409) (5,0.9408)
				(6,0.9411) (7,0.9448) (8,0.9413) (9,0.9448) (10,0.9411)
				(11,0.9411) (12,0.9334) (13,0.9413) (14,0.9411) (15,0.9411)
				(16,0.9411) (17,0.9411) (18,0.9411) (19,0.9411) (20,0.9411)
			};
			
			% OPT model
			\addplot[
			color=color2,
			mark=triangle*,
			mark options={solid,fill=color2}
			] coordinates {
				(1,0.8703) (2,0.8848) (3,0.8905) (4,0.8876) (5,0.8872)
				(6,0.8876) (7,0.8911) (8,0.8802) (9,0.8911) (10,0.891)
				(11,0.8911) (12,0.8945) (13,0.891) (14,0.951) (15,0.931)
				(16,0.9558) (17,0.891) (18,0.9558) (19,0.9758) (20,0.8911)
			};
			
			% Phi model
			\addplot[
			color=color3,
			mark=pentagon*,
			mark options={solid,fill=color3}
			] coordinates {
				(1,0.8799) (2,0.8988) (3,0.891) (4,0.8892) (5,0.9517)
				(6,0.9317) (7,0.8917) (8,0.8858) (9,0.9358) (10,0.9059)
				(11,0.956) (12,0.8858) (13,0.8917) (14,0.8917) (15,0.9217)
				(16,0.896) (17,0.8917) (18,0.9483) (19,0.9594) (20,0.8858)
			};
			
			% TinyLlama model
			\addplot[
			color=color4,
			mark=diamond*,
			mark options={solid,fill=color4}
			] coordinates {
				(1,0.8762) (2,0.8847) (3,0.8862) (4,0.8862) (5,0.88)
				(6,0.884) (7,0.8807) (8,0.888) (9,0.884) (10,0.884)
				(11,0.884) (12,0.8873) (13,0.884) (14,0.8807) (15,0.8932)
				(16,0.884) (17,0.8807) (18,0.8898) (19,0.884) (20,0.884)
			};
			
			\legend{
				DeepSeek-R1-Distill-Qwen-1.5B,
				facebook/opt-1.3b,
				microsoft/phi-1.5,
				TinyLlama-1.1B
			}
		\end{axis}
	\end{tikzpicture}
	\caption{ROC AUC values across 20 communication rounds for four models.}
	\label{fig:main_comparison}
\end{figure*}

\begin{table*}[h]
	\centering
	\caption{Statistical Metrics  on BGL Dataset}
	\label{tab:performance_metrics}
	\begin{tabular}{lccccc}
		\toprule
		\textbf{Model} & \textbf{Mean} & \textbf{Std Dev} & \textbf{Min} & \textbf{Max} & \textbf{Range} \\
		\midrule
		DeepSeek-R1-Distill-Qwen-1.5B & 0.9411 & 0.0026 & 0.9334 & 0.9448 & 0.0114 \\
		facebook/opt-1.3b & 0.9082 & 0.0296 & 0.8703 & 0.9758 & 0.1055  \\
		microsoft/phi-1.5 & 0.9089 & 0.0272 & 0.8799 & 0.9594 & 0.0795  \\
		TinyLlama-1.1B & 0.8843 & 0.0036 & 0.8762 & 0.8932 & 0.0170  \\
		\bottomrule
	\end{tabular}
\end{table*}

Figure \ref{fig:main_comparison} and Table\ref{tab:best_performance1} present the ROC-AUC values for FlogTinyLLM on the BGL dataset across 20 communication rounds. DeepSeek-R1 achieves values ranging from 0.9334 to 0.9448 with a mean of 0.9411. The Opt-1.3b model exhibits values ranging from 0.8703 to 0.9758, with a mean of 0.9082 and a peak single-round performance at round 19. The Phi-1.5 model shows values between 0.8799 and 0.9594, with a mean 0.9089, with a peak at round 19 (0.9594). TinyLlama-1.1B displays values from 0.8762 to 0.8932 with a mean of 0.8843. DeepSeek-R1 demonstrates the lowest variation, while opt-1.3b shows the largest.

\begin{table*}[h]
	\centering
	\caption{FlogTinyLLM Best Performances Across All Rounds on BGL Dataset}
	\label{tab:best_performance1}
	\begin{tabular}{@{}lcccccc@{}}
		\toprule
		\textbf{Model} & \textbf{Best} & \textbf{Best} & \textbf{Best} & \textbf{Best} & \textbf{Best} & \textbf{Round} \\ 
		& \textbf{Acc.} & \textbf{F1} & \textbf{Prec.} & \textbf{Rec.} & \textbf{AUC} & \textbf{Round} \\ 
		\midrule
		DeepSeek-R1-Distill-Qwen-1.5B & 0.9681 & 0.8922 & 1.0000 & 0.8142 & 0.9448 & 7--9 \\
		OPT-1.3B & 0.9332 & 0.9023 & 1.0000 & 0.8903 & 0.9758 & 19 \\
		TinyLlama-1.1B & 0.9492 & 0.8168 & 1.0000 & 0.6903 & 0.8932 & 7, 15--18 \\
		Phi-1.5 & 0.9492 & 0.9168 & 1.0000 & 0.8903 & 0.9594 & 18 \\
		\bottomrule
	\end{tabular}
\end{table*}

\begin{table*}[h]
	\centering
	\caption{Average Performance Metrics Across 20 Federated Learning Rounds}
	\label{tab:average_performance}
	\begin{tabular}{@{}lccccc@{}}
		\toprule
		\textbf{Model} & \textbf{Accuracy} & \textbf{F1} & \textbf{Precision} & \textbf{Recall} & \textbf{AUC} \\ 
		\midrule
		DeepSeek-R1-Distill-Qwen-1.5B & \textbf{0.9569} & \textbf{0.8609} & 0.9268 & \textbf{0.8057} & \textbf{0.9410} \\
		OPT-1.3B & 0.9324 & 0.8051 & 0.8737 & 0.7730 & 0.9045 \\
		TinyLlama-1.1B & 0.9368 & 0.7822 & 0.9041 & 0.6899 & 0.8843 \\
		Phi-1.5 & 0.9448 & 0.8260 & \textbf{0.9666} & 0.7569 & 0.9095 \\
		\bottomrule
	\end{tabular}
\end{table*}

Table \ref{tab:average_performance} show the mean performance metrics of FlogTinyLLM on the BGL Data set over 20 rounds. \textbf{DeepSeek-R1-Distill-Qwen-1.5B} ranks first in 4 of 5 metrics (Accuracy, F1, Recall, AUC), showing balanced and consistently high performance across all evaluation criteria.
\textbf{Phi-1.5} excels in precision (0.9666), the highest among all models, and maintains competitive accuracy and AUC. However, its recall (0.7569) ranks third.
\textbf{TinyLlama-1.1B} has the lowest recall (0.6899).
\textbf{OPT-1.3B} shows moderate performance across all metrics, ranking last in precision (0.8737) but second in recall (0.7730). All models maintain accuracy above 0.93, showing effective federated learning training across the 20 rounds.

\subsection{Performance Analysis of FlogTinyLLM and LogTinyLLM on the BGL Dataset}
% Table 1: Side-by-side comparison
\begin{table*}[h]
	\centering
	\caption{Performance Comparison: FlogTinyLLM vs. LogTinyLLM Training on BGL Dataset}
	\label{tab:federated_vs_centralized}
	\resizebox{\textwidth}{!}{%
		\begin{tabular}{@{}lcccccccc@{}}
			\toprule
			\multirow{2}{*}{\textbf{Model}} & \multicolumn{4}{c}{\textbf{FlogTinyLLM}} & \multicolumn{4}{c}{\textbf{LogTinyLLM}} \\
			\cmidrule(lr){2-5} \cmidrule(lr){6-9}
			& \textbf{Acc.} & \textbf{F1} & \textbf{Prec.} & \textbf{Rec.} & \textbf{Acc.} & \textbf{F1} & \textbf{Prec.} & \textbf{Rec.} \\
			\midrule
			DeepSeek-R1-Distill-Qwen-1.5B & 0.9569 & 0.8609 & 0.9268 & 0.8057 & 0.9911 & 0.9912 & 0.9913 & 0.9849 \\
			OPT-1.3B                      & 0.9324 & 0.8051 & 0.8737 & 0.7730 & 0.9911 & 0.9913 & 0.9913 & 0.9693 \\
			TinyLlama-1.1B                & 0.9368 & 0.7822 & 0.9041 & 0.6899 & 0.9916 & 0.9917 & 0.9917 & 0.9856 \\
			Phi-1.5                       & 0.9448 & 0.8260 & 0.9666 & 0.7569 & 0.9919 & 0.9919 & 0.9920 & 0.9862 \\
			\bottomrule
		\end{tabular}%
	}
\end{table*}

\begin{table*}[h]
	\centering
	\caption{Performance Gap: LogTinyLLM vs. FlogTinyLLM (Difference)}
	\label{tab:performance_gap}
	\begin{tabular}{@{}lcccc@{}}
		\toprule
		\textbf{Model} & \textbf{$\Delta$ Acc.} & \textbf{$\Delta$ F1} & \textbf{$\Delta$ Prec.} & \textbf{$\Delta$ Rec.} \\
		\midrule
		DeepSeek-R1-Distill-Qwen-1.5B & +0.0342 & +0.1303 & +0.0645 & +0.1792 \\
		OPT-1.3B                      & +0.0587 & +0.1862 & +0.1176 & +0.1963 \\
		TinyLlama-1.1B                & +0.0548 & +0.2095 & +0.0876 & +0.2957 \\
		Phi-1.5                       & +0.0471 & +0.1659 & +0.0254 & +0.2293 \\
		\midrule
		\textbf{Mean Gap}             & \textbf{+0.0487} & \textbf{+0.1730} & \textbf{+0.0738} & \textbf{+0.2251} \\
		\bottomrule
	\end{tabular}
\end{table*}

The results presented in Table~\ref{tab:federated_vs_centralized} indicate that LogTinyLLM  outperforms the FlogTinyLLM across the four models in terms of accuracy, F1-score, precision, and recall on the BGL dataset. Specifically, LogTinyLLM achieves accuracies between 0.9911 and 0.9919, F1-Scores between 0.9912 and 0.9919, precisions between 0.9913 and 0.9920, and recalls between 0.9693 and 0.9862, with loss values ranging from 0.0307 to 0.0335. In comparison, FlogTinyLLM demonstrates lower performance, with accuracy ranging from 0.9324 to 0.9569, F1-score from 0.7822 to 0.8609, precision from 0.8737 to 0.9666, recall from 0.6899 to 0.8057, and AUC from 0.8843 to 0.9410. Although the FlogTiny architecture performs well on the BGL dataset, its results do not match those of LogTinyLLM. Table~\ref{tab:performance_gap} provides a quantitative assessment of these differences, with positive deltas reflecting the superior performance of LogTinyLLM. The mean differences are +0.0487 for accuracy, +0.1730 for F1-score, +0.0738 for precision, and +0.2251 for recall. The most substantial individual gap is observed in recall for TinyLlama-1.1B (+0.2957) and in F1-score for TinyLlama-1.1B (+0.2095), while the smallest precision gap is noted for Phi-1.5 (+0.0254).

\definecolor{deepseek}{RGB}{44,62,80}      % Midnight blue-gray
\definecolor{opt}{RGB}{149,165,166}        % Silver-gray
\definecolor{tinyllama}{RGB}{52,73,94}     % Wet asphalt
\definecolor{phi}{RGB}{127,140,141}        % Concrete gray

% Accent colors for federated vs centralized
\definecolor{federatedblue}{RGB}{52,73,94}     % Deep slate blue
\definecolor{centralizedgold}{RGB}{211,84,0}   % Burnt orange/copper
\definecolor{gridgray}{RGB}{236,240,241}       % Very light gray for grids

\begin{figure}[h]
	\centering
	\begin{tikzpicture}
		\begin{axis}[
			xbar,
			bar width=0.20cm,
			width=0.8\textwidth,
			height=8cm,
			xlabel={Performance Gap},
			symbolic y coords={DeepSeek-R1, OPT-1.3B, TinyLlama-1.1B, Phi-1.5},
			% ytick=data,
			% y tick label style={font=\large},
			xtick={0, 0.05, 0.10, 0.15, 0.20, 0.25, 0.30},
			xmin=0,
			xmax=0.32,
			legend style={
				at={(0.98,0.02)},
				anchor=south east,
				legend columns=2,
				font=\normalsize,
				draw=black!15,
				line width=0.5pt,
				fill=white,
				fill opacity=0.92,
				text opacity=1,
				cells={anchor=west},
				rounded corners=0.9pt,
				inner sep=6pt
			},
			xmajorgrids=true,
			grid style={dashed, draw=gridgray, line width=0.5pt},
			enlarge y limits=0.20,
			axis line style={line width=0pt},
			% major tick style={line width=1.2pt},
			nodes near coords,
			nodes near coords style={
				% font=\footnotesize\bfseries,
				/pgf/number format/fixed,
				/pgf/number format/precision=3,
				anchor=west,
				xshift=0.8pt
			},
			]
			
			% Accuracy gaps
			\addplot[
			fill=deepseek!75,
			draw=deepseek,
			line width=1.5pt
			] coordinates {
				(0.0342,DeepSeek-R1) (0.0587,OPT-1.3B) (0.0548,TinyLlama-1.1B) (0.0471,Phi-1.5)
			};
			
			% F1-Score gaps
			\addplot[
			fill=opt!80,
			draw=opt!120,
			line width=1.5pt,
			postaction={pattern=dots, pattern color=opt!50}
			] coordinates {
				(0.1303,DeepSeek-R1) (0.1862,OPT-1.3B) (0.2095,TinyLlama-1.1B) (0.1659,Phi-1.5)
			};
			
			% Precision gaps
			\addplot[
			fill=tinyllama!75,
			draw=tinyllama,
			line width=1.5pt,
			postaction={pattern=horizontal lines, pattern color=tinyllama!40}
			] coordinates {
				(0.0645,DeepSeek-R1) (0.1176,OPT-1.3B) (0.0876,TinyLlama-1.1B) (0.0254,Phi-1.5)
			};
			
			% Recall gaps
			\addplot[
			fill=phi!85,
			draw=phi!130,
			line width=1.5pt,
			postaction={pattern=vertical lines, pattern color=phi!50}
			] coordinates {
				(0.1792,DeepSeek-R1) (0.1963,OPT-1.3B) (0.2957,TinyLlama-1.1B) (0.2293,Phi-1.5)
			};
			
			\legend{Accuracy, F1-Score, Precision, Recall}
		\end{axis}
	\end{tikzpicture}
	\caption{Performance gap analysis with horizontal bars showing gaps by model across metrics}
	\label{fig:performance_gap_horizontal}
\end{figure}
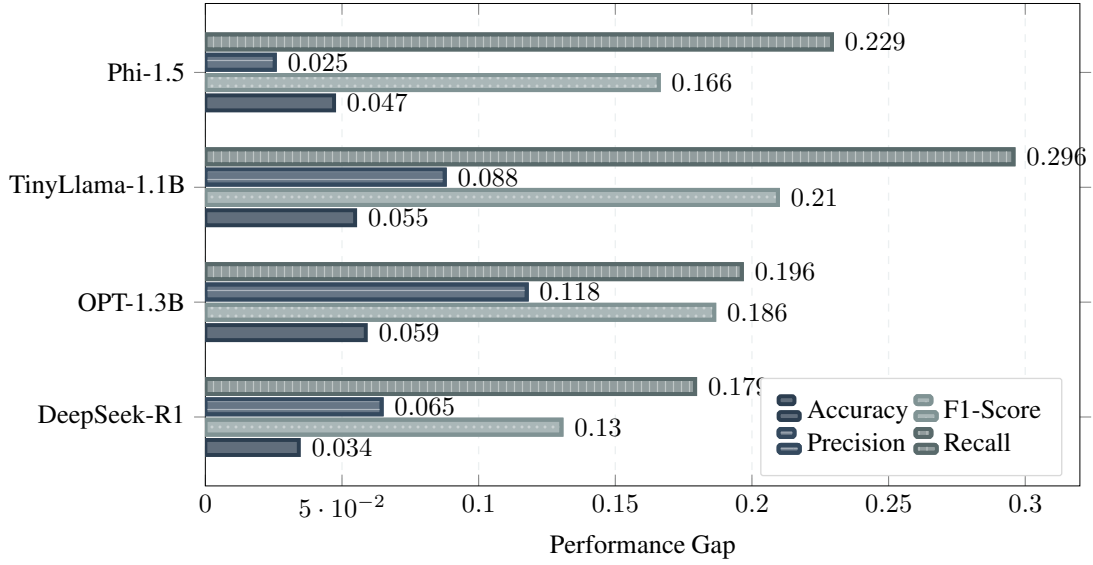

The Fig \ref{fig:performance_gap_horizontal} is a horizontal bar chart. It compares performance gaps (LogTinyLLM minus FlogTinyLLM) across four metrics: Accuracy, F1-Score, Precision, and Recall. The x-axis shows the gap values from 0 to 0.32. The y-axis lists the models. Each metric uses bars with distinct fills and patterns.

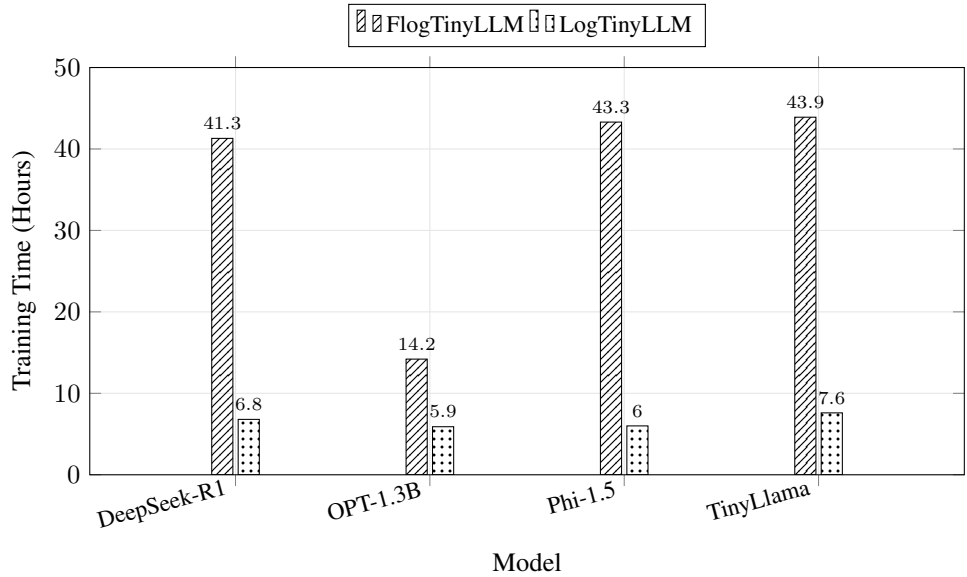
\begin{figure*}[t]
\centering
\begin{tikzpicture}
\begin{axis}[
    ybar,
    bar width=8pt,
    width=0.8\textwidth,
    height=7cm,
    ylabel={Training Time (Hours)},
    xlabel={Model},
    symbolic x coords={DeepSeekR1, OPT13B, Phi15, TinyLlama},
    xtick=data,
    xticklabels={DeepSeek-R1, OPT-1.3B, Phi-1.5, TinyLlama},
    x tick label style={rotate=15, anchor=east, font=\small},
    ymin=0,
    ymax=50,
    enlarge x limits=0.25,
    nodes near coords,
    nodes near coords style={font=\scriptsize},
    grid=major,
    grid style={gray!20},
    legend style={
        at={(0.5,1.05)},
        anchor=south,
        legend columns=2,
        font=\footnotesize
    }
]

% Federated
\addplot[
    fill=fedcolor,
    draw=black,
    pattern=north east lines
] coordinates {
    (DeepSeekR1,41.3)
    (OPT13B,14.2)
    (Phi15,43.3)
    (TinyLlama,43.9)
};

% Centralized
\addplot[
    fill=centcolor,
    draw=black,
    pattern=dots
] coordinates {
    (DeepSeekR1,6.8)
    (OPT13B,5.9)
    (Phi15,6.0)
    (TinyLlama,7.6)
};

\legend{FlogTinyLLM, LogTinyLLM}

\end{axis}
\end{tikzpicture}

\caption{Comparison of training time between FlogTinyLLM and LogTinyLLM across the four model architectures on the BGL dataset.}
\label{fig:training_time_bgl}
\end{figure*}

Figure \ref{fig:training_time_bgl} shows that OPT-1.3B in the FlogTinyLLM architecture achieves the fastest federated training on the BGL dataset, completing 20 rounds in 14.18 hours. This makes it 3.09$\times$ faster than TinyLlama-1.1B (43.85 hours), despite having more parameters (714.1M vs. 556.4M). DeepSeek-R1-Distill-Qwen-1.5B follows with 41.29 hours, while Phi-1.5 needs 43.31 hours. For LogTinyLLM, training time decreases across all models. OPT-1.3B remains most efficient, finishing centralized training in 5.92 hours, followed by Phi-1.5 at 6.03 hours, DeepSeek-R1 at 6.83 hours, and TinyLlama-1.1B at 7.60 hours. The tight cluster of 1.68 hours shows similar computational needs across architectures with centralized data.

\subsubsection{Performance  Comparison with Federated Baseline Models on the BGL Dataset}

\definecolor{charcoal}{HTML}{3D3D3D}
\definecolor{slategray}{HTML}{6B7B8D}
\definecolor{warmtaupe}{HTML}{A89F91}
\definecolor{mutedteal}{HTML}{5F8A8B}
\definecolor{warmgray}{HTML}{D6CFC7}

\begin{table}[h]
	\caption{Performance Comparison of Federated Learning Models on the BGL Dataset}
	\label{tab:model_comparison542}
	\centering
	\begin{tabular}{lccc}
		\toprule
		\textbf{Model} & \textbf{Precision} & \textbf{Recall} & \textbf{F1-Score} \\
		\midrule
		Federated LogBERT              & 0.7574          & \textbf{0.9625} & 0.8477          \\
		Federated LogGPT               & 0.9211          & 0.7482          & 0.7562          \\
		LogDeep(FL) & \textbf{0.9688} & 0.8611          & \textbf{0.9118} \\
		FlogTinyLLM: OPT 1.3B         & 0.8737          & 0.7730          & 0.8051          \\
		FlogTinyLLM: DeepSeek          & 0.9268          & 0.8057          & 0.8609          \\
		FlogTinyLLM: TinyLLama         & 0.9041          & 0.6899          & 0.7822          \\
		FlogTinyLLM: Phi 1.5           & 0.9666          & 0.7569          & 0.8260          \\
		\bottomrule
		\multicolumn{4}{l}{\footnotesize Bold values indicate the best performance for each metric.}
	\end{tabular}
\end{table}

Table \ref{tab:model_comparison542} compares federated learning models for log anomaly detection on the BGL dataset. LogDeep(FL) achieves the highest precision (0.9688) and F1-score (0.9118), showing a strong balance between precision and recall. Federated LogBERT delivers the best recall (0.9625), though at the expense of lower precision. Among the FlogTinyLLM variants, DeepSeek attains the best F1-score (0.8609), while Phi 1.5 reaches the highest precision (0.9666). Federated LogGPT has the lowest F1 Score (0.7562), showing a weaker precision-recall trade-off.

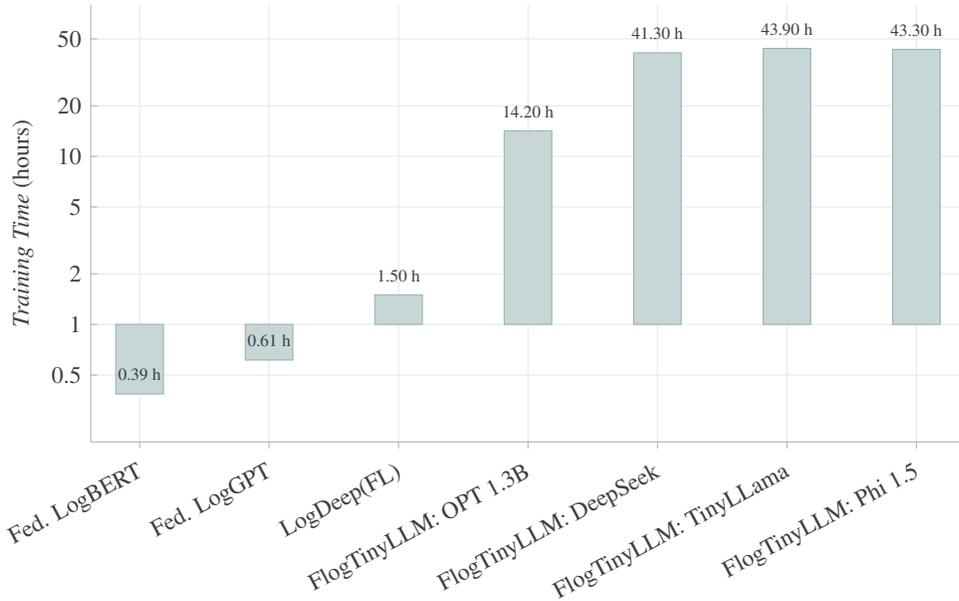
\begin{figure}[hbt]
	\centering
	\begin{tikzpicture}
		\begin{axis}[
			width=0.8\textwidth,
			height=7.4cm,
			ybar,
			bar width=18pt,
			enlarge x limits={abs=0.65cm},
			%
			% ?? Y-axis (log scale, hours) ??
			ylabel={\textit{Training Time} (hours)},
			ylabel style={font=\small, color=charcoal},
			ymode=log,
			log basis y=10,
			ymin=0.2,
			ymax=80,
			ytick={0.5,1,2,5,10,20,50},
			yticklabels={0.5,1,2,5,10,20,50},
			yticklabel style={font=\footnotesize, color=charcoal},
			%
			% ?? X-axis (models) ??
			symbolic x coords={
				Fed. LogBERT,
				Fed. LogGPT,
				LogDeep(FL),
				FlogTinyLLM: OPT 1.3B,
				FlogTinyLLM: DeepSeek,
				FlogTinyLLM: TinyLLama,
				FlogTinyLLM: Phi 1.5
			},
			xtick=data,
			xticklabel style={
				font=\footnotesize,
				color=charcoal,
				rotate=28,
				anchor=north east,
			},
			%
			% ?? Grid & frame ??
			grid=major,
			grid style={thin, warmgray!50},
			axis line style={charcoal!40, thin},
			every tick/.style={charcoal!40},
			major tick length=2.5pt,
			axis x line*=bottom,
			axis y line*=left,
			%
			% ?? Value labels on bars ??
			nodes near coords,
			every node near coord/.append style={
				font=\fontsize{6.5}{7.5}\selectfont,
				color=charcoal,
				anchor=south,
				yshift=1.5pt,
			},
			point meta=explicit symbolic,
			clip=false,
			]
			
			\addplot[
			fill=mutedteal!35,
			draw=mutedteal!65,
			fill opacity=0.92,
			] coordinates {
				(Fed. LogBERT,           0.3856)  [0.39 h]
				(Fed. LogGPT,            0.6138)  [0.61 h]
				(LogDeep(FL),          1.5000)  [1.50 h]
				(FlogTinyLLM: OPT 1.3B, 14.2000) [14.20 h]
				(FlogTinyLLM: DeepSeek,  41.3000) [41.30 h]
				(FlogTinyLLM: TinyLLama, 43.9000) [43.90 h]
				(FlogTinyLLM: Phi 1.5,   43.3000) [43.30 h]
			};
			
		\end{axis}
	\end{tikzpicture}
	\caption{Training time efficiency of federated log anomaly detection models on the BGL dataset (log scale). Federated LogBERT completes training in 1388.19\,seconds ($\approx$\,0.39\,hours), while the FlogTinyLLM variants require between 14.2 and 43.9\,hours due to the computational overhead of fine-tuning large language models in a federated setting.}
	\label{fig:training_time}
\end{figure}

Figure~\ref{fig:training_time} presents a bar chart of training times in hours for federated log anomaly detection models on the BGL dataset, with a logarithmic y-scale (base 10) from 0.2 to 80 and y-ticks at 0.5, 1, 2, 5, 10, 20, 50.

Federated LogBERT completes training in 1388.19 seconds ($\approx$0.39 hours), while the FlogTinyLLM variants require between 14.2 and 43.9 hours due to the computing overhead of fine-tuning large language models in a federated setting.

\begin{table}[h]
	\centering
	\renewcommand{\arraystretch}{1.35}
	\caption{Performance Comparison of Federated Models on the Thunderbird Dataset}
	\label{tab:comparison123}
	\vspace{0.3em}
	\begin{tabular}{l c c c}
		\toprule
		\textbf{Model} & \textbf{Recall} & \textbf{Precision} & \textbf{F1-Score} \\
		\midrule
		Federated LogBERT                  & 0.9625          & 0.7574             & 0.8477            \\
		LogDeep(FL) & \textbf{1.00}   & 0.7935             & 0.8191            \\
		Federated LogGPT                   & 0.9924          & 0.9629             & 0.9774            \\
		FlogTinyLLM: Phi-1.5               & 0.9828          & 0.9994             & 0.9908            \\
		FlogTinyLLM: DeepSeek-R1           & 0.9615          & \textbf{0.9996}    & 0.9797            \\
		FlogTinyLLM: OPT-1.3               & 0.9879          & \textbf{0.9996}    & \textbf{0.9935}   \\
		FlogTinyLLM: TinyLlama             & 0.9354          & 0.9991             & 0.9660            \\
		\bottomrule
	\end{tabular}
\end{table}

Table \ref{tab:comparison123} presents a comparison of federated learning-based log anomaly detection models evaluated on the Thunderbird dataset. The best result for each metric is highlighted in the table.

The FlogTinyLLM variants consistently outperform traditional federated approaches (LogBERT, LogDeep(FL), LogGPT) in F1-Score on the Thunderbird dataset, showing the effectiveness of tiny LLMs for federated log anomaly detection.

\begin{figure}[]
	\centering
	\includegraphics[width=0.85\textwidth]{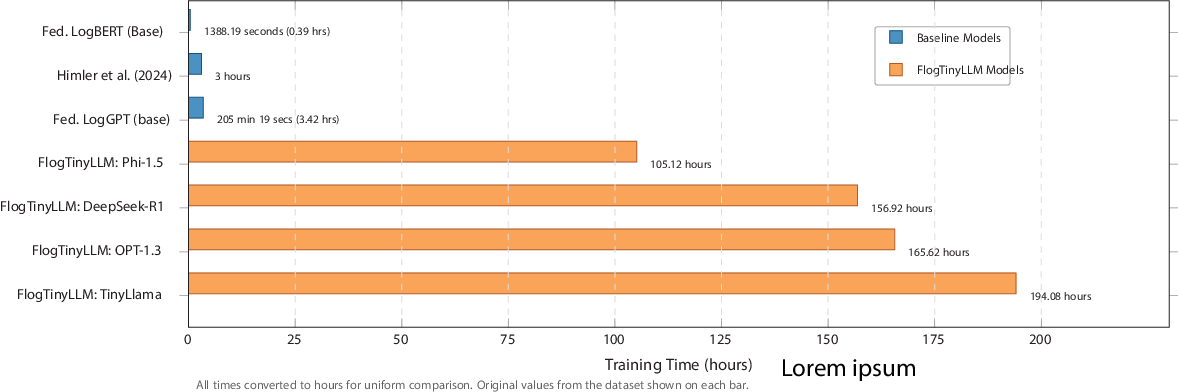}
	\caption{Training time comparison of federated models on the Thunderbird dataset.}
	\label{fig:training_time123}
\end{figure}

\section{Discussion}

Comparing FlogTinyLLM and LogTinyLLM across both  Thunderbird and BGL datasets shows that FlogTinyLLM provides strong local training on distributed clients while preserving data privacy. This approach produces results comparable to LogTinyLLM's centralised training on both datasets, supporting the feasibility of privacy-preserving and practical learning. On the Thunderbird dataset, FlogTinyLLM achieves near-centralized performance, with accuracy scores of 0.9876 for OPT-1.3B and 0.9825 for Phi-1.5. Precision in the federated setup remains high, slightly exceeding that of the centralized setup. Recall and F1-scores are also competitive, demonstrating FlogTinyLLM's effectiveness for Thunderbird. On the BGL dataset, FlogTinyLLM's accuracy ranges from 0.9324 to 0.9569 and F1-scores from 0.7822 to 0.8609. While the FlogTinyLLM trails centralized LogTinyLLM on BGL, the gap across models is modest. Training times are longer for FlogTinyLLM due to privacy measures on both datasets, with OPT-1.3B notable for faster federated training and near-centralized accuracy. Overall, FlogTinyLLM performs well on both datasets with reasonable increases in computational cost. Comparing FlogtinyLLM family with benchmark models: Federated LogBERT, Federated LogGPT, and LogDeep(FL)  on the Thunderbird dataset from Table~\ref{tab:comparison123}. It can be seen that within the FlogTinyLLM family, all variants obtain high Precision ($\geq$0.9991). However, they differ slightly in Recall. OPT-1.3 and Phi-1.5 achieve the highest Recall values (0.9879 and 0.9828). TinyLlama records the lowest Recall in this group (0.9354). DeepSeek-R1 falls in between at 0.9615. These results suggest that among the FlogTinyLLM family, OPT-1.3 provides the best anomaly detection capability in federated settings on the Thunderbird dataset. Phi-1.5 offers a close second. It can also be seen from Table ~\ref{tab:comparison123} that the FlogTinyLLM family outperformed all the benchmark models on precision and F1-Score, except on recall, where LogDeep(FL) leads with perfect recall. Figure ~\ref{fig:training_time123} shows a difference in training time between baselines and FlogTinyLLM models. The three baselines Federated LogBERT (0.39 hrs), LogDeep(FL) (3 hrs), and Federated LogGPT (3.42 hrs) train in under 4 hours. In contrast, FlogTinyLLM variants require more time, from 105.12 hours (Phi-1.5) to 194.08 hours (TinyLlama), about 30 to 50 times longer than baselines. Comparing the FlogTinyLLM family with the benchmark models Federated LogBERT, Federated LogGPT, and LogDeep(FL), Table ~\ref{tab:model_comparison542} shows their performance measures on the BGL dataset. Regarding precision, LogDeep(FL) achieves the highest (0.9688), with FlogTinyLLM: Phi-1.5 only slightly lower (0.9666). In contrast, Federated LogBERT obtains the lowest precision (0.7574). Other FlogTinyLLM variants fall between 0.8737 (OPT-1.3B) and 0.9268 (DeepSeek). This demonstrates that, in terms of precision, FlogTinyLLM variants generally outperform Federated LogBERT but not FL LogDeep on the BGL dataset. For recall metrics on the BGL dataset, Federated LogBERT leads at 0.9625, excelling at true-positive identification. FlogTinyLLM: TinyLlama has the lowest recall (0.6899).  Regarding F1-scores: LogDeep(FL) lead at 0.9118, with FlogTinyLLM: DeepSeek (0.8609) and Federated LogBERT (0.8477) following. The lowest F1-scores are seen in Federated LogGPT (0.7562) and FlogTinyLLM: TinyLLama (0.7822). Notably, some FlogTinyLLM variants, such as DeepSeek, remain competitive, offering effective trade-offs. It can also be seen from Figure ~\ref{fig:training_time} that the FlogTinyLLM family took more time to train due to the use of LLMs

\section{Conclusion}

In this work, we studied log anomaly detection in settings where data cannot be centrally shared. We proposed DP-FLogTinyLLM, a federated framework that integrates differential privacy with parameter-efficient LLM fine tuning. Empirical results on the Thunderbird and BGL datasets show that FlogTinyLLM performs close to its centralized counterpart in terms of accuracy, precision, and F1-score. The differences across models are small, suggesting that the federated setup is able to retain most of the predictive performance while adding privacy protection. On Thunderbird, the results are nearly identical to the centralized setting, while on BGL the model remains strong across all metrics with only slight variation in recall. Also, FlogTinyLLM outperforms benchmark federated models such as Federated LogBERT, Federated LogGPT, and LogDeep(FL) on the Thunderbird dataset, while remaining competitive on BGL. Moving to a federated and privacy-preserving setup does introduce additional computational cost. However, this overhead appears to be a reasonable trade-off given that sensitive log data never leaves local systems.

\clearpage

\bibliographystyle{unsrtnat}
\bibliography{references}  %%% Uncomment this line and comment

\end{document}